\documentclass[prx,amsmath,amssymb, notitlepage, twocolumn,floatfix,nofootinbib,superscriptaddress, longbibliography]{revtex4-1}
\usepackage{amsmath}
\usepackage{mathtools} 
\usepackage{tabularx,graphicx}
\usepackage{epstopdf}
\usepackage{graphicx}
\usepackage{enumerate}
\usepackage{spectralsequences} 
\usepackage{latexsym}
\usepackage{amssymb}
\usepackage{tikz-cd}
\usepackage{amsmath}
\usepackage{color, colortbl}
\usepackage{psfrag}
\usepackage{bbm}
\usepackage{bm}
\usepackage{titlesec}
\usepackage{dsfont}
\usepackage{feynmp}
\usepackage{slashed}
\usepackage{multirow}
\usepackage{framed}
\usepackage{braket}
\newcommand{\beq}{\begin{eqnarray}}
\newcommand{\eeq}{\end{eqnarray}}

\newcommand{\Z}{\mathbb{Z}}

\newcommand{\bs}{\boldsymbol}

\newcommand{\bsp}{\begin{split}}
\newcommand{\esp}{\end{split}}

\newcommand{\D}{\mathcal{D}}

\usepackage{color}
\definecolor{darkblue}{rgb}{0.,0.,0.4}
\definecolor{darkred}{rgb}{0.5,0.,0.}
\definecolor{BlueViolet}{RGB}{138,43,226}
\definecolor{SkyBlue}{RGB}{30,144,255}
\definecolor{DarkGreen}{RGB}{0,100,0}
\usepackage[pdftex,colorlinks=true,linkcolor=darkblue,citecolor=blue,urlcolor=darkred]{hyperref}
\usepackage[normalem]{ulem}

\makeatletter
\newsavebox{\@brx}
\newcommand{\llangle}[1][]{\savebox{\@brx}{\(\m@th{#1\langle}\)}%
  \mathopen{\copy\@brx\kern-0.5\wd\@brx\usebox{\@brx}}}
\newcommand{\rrangle}[1][]{\savebox{\@brx}{\(\m@th{#1\rangle}\)}%
  \mathclose{\copy\@brx\kern-0.5\wd\@brx\usebox{\@brx}}}
\makeatother
\usepackage{amsthm}
\theoremstyle{plain}
\newtheorem*{theorem*}{Theorem}

\renewcommand{\vec}[1]{\bm{#1}}

\begin{document}
\title{Fidelity Strange Correlators for Average Symmetry-Protected Topological Phases}
\author{Jian-Hao Zhang}
\affiliation{Department of Physics, The Pennsylvania State University, University Park, Pennsylvania 16802, USA}
\affiliation{Department of Physics and Center for Theory of Quantum Matter, University of Colorado, Boulder, Colorado 80309, USA}
\author{Yang Qi}
\email{qiyang@fudan.edu.cn}
\affiliation{State Key Laboratory of Surface Physics, Fudan University, Shanghai 200433, China}
\affiliation{Center for Field Theory and Particle Physics, Department of Physics, Fudan University, Shanghai 200433, China}
\affiliation{Collaborative Innovation Center of Advanced Microstructures, Nanjing 210093, China}
\author{Zhen Bi}
\email{zjb5184@psu.edu}
\affiliation{Department of Physics, The Pennsylvania State University, University Park, Pennsylvania 16802, USA}

\begin{abstract}
Topological phases in open quantum systems have become an exciting area of research, driven by the rising importance of noisy intermediate-scale quantum platforms. The concept of Average Symmetry-Protected Topological (ASPT) phases extends symmetry-protected topological phases to quantum systems affected by disorder or decoherence. To identify nontrivial ASPT states, we introduce a ``fidelity strange correlator" (FSC), which operates directly on a single bulk density matrix without boundaries. For nontrivial ASPT phases in one and two dimensions, we show that the FSC reveals long-range or power-law behavior. In several two-dimensional examples, we establish a link between the FSC and certain non-local correlation functions in statistical loop models with quantum corrections, enabling us to derive exact scaling exponents for the FSC. In addition, we discuss methods to measure the FSC using classical shadow tomography. Our work lays the groundwork for identifying these intriguing topological phases of matter in open quantum systems, both numerically and experimentally. 
\end{abstract}

\maketitle

\section{Introduction} 

Symmetry-protected topological (SPT) phases host nontrivial short-range entanglement (SRE) which cannot be destroyed in the presence of symmetries \cite{ZCGu2009,chen11a, XieChenScience, cohomology, Senthil_2015, Lu12, general2, Kapustin2017,2DFSPT, LevinGu, braiding, Ashvin2013, ChongWang2013, Senthil2013, Lukasz2013, ChongWang2014, TCI, Fu2012, ITCI, reduction, building, correspondence,230, rotation, dihedral, wallpaper,indicator1, JHZhang2022, Zhang_2022, Kruthoff_2017, Slager_2012}. Recently, it is shown that symmetry-protected SRE can still prevail even if part of the protecting symmetry is broken locally by decoherence or quenched disorder but restored upon ensemble averaging \cite{MaWangASPT, Ringel_2012, Mong_2012, Fulga_2014, Chou_2018, Chou_2021,ma2025topological}, which defines a new class of SPT phases dubbed average SPT (ASPT). ASPTs extend the concept of SPT phases to include systems where symmetry is statistically preserved rather than strictly maintained, opening up new possibilities for exploring topological phenomena in disordered systems. Moreover, the concept of ASPTs naturally extends to mixed states, which arise in systems coupled to an environment, where quantum decoherence plays a significant role. 
Given that symmetry-breaking disorders or quantum decoherence are practically unavoidable in experimental settings, investigating the properties of ASPTs holds both theoretical importance and practical relevance. Understanding ASPTs not only enriches our comprehension of topological phases in real-world systems but also provides guidance for designing and interpreting experiments in noisy or disordered quantum environments.

The nontrivial features of SPT phases manifest on the physical boundaries which often show symmetry-protected gapless spectrum, while bulk correlation functions of local observables all decay exponentially with distance due to the spectral gap. This bulk-boundary dichotomy makes the detection of SPT phases particularly challenging when only access to a single bulk wavefunction, without physical boundaries, is available. One powerful tool is the \textit{strange correlator} \cite{Strange} defined for a given wavefunction $|\Psi\rangle$ as 
\begin{align}
\label{strange pure}
C(r,r')=\frac{\langle\Psi|O(r)O(r')|\Psi_0\rangle}{\langle\Psi|\Psi_0\rangle}.
\end{align}
Here $|\Psi_0\rangle$ is a symmetric trivial state serving as a reference, and $O$'s are certain local operators. It has been demonstrated that, given a nontrivial SPT wavefunction, the strange correlator is generically \textit{long-range} or \textit{power-law} in the long-distance limit\cite{Strange,Paganelli_2022, Scaffidi_2016}. The success of the strange correlator can be understood through a Wick rotation perspective, where the strange correlator is mapped to the correlation function at the boundary of the SPT phase. This boundary correlation function is guaranteed to be nontrivial due to the presence of anomalous surface states inherent to the SPT phase. The strange correlator has been successfully applied to identify nontrivial SPT wavefunctions in numerical simulations \cite{Wu_2015, Vanhove_2018, Meng_2022, he2016bona, wierschem2014quenching, vanhove2022topological, wierschem2016detection, wierschem2014strange}. 

This work extends the concept of the strange correlator to ASPT states by introducing a novel tool, the fidelity strange correlator (FSC), which leverages the concept of fidelity from quantum information theory. We highlight that the fidelity between two density matrices is a natural extension of wavefunction overlap in mixed states, which allows us to define a basis-independent form of strange correlator for ASPTs. 
We also explore potential methods for measuring the FSC using classical shadow tomography, a powerful and efficient technique for estimating properties of quantum states. By combining the FSC framework with classical shadow methods, our work provides a concrete pathway for identifying and characterizing ASPT phases in noisy or decoherent quantum systems, bridging the gap between theoretical proposals and experimental realizations.

For the rest of the paper, we first briefly review the notion of ASPTs in the introduction section. In Sec II, we define the FSC for ASPTs. Then we showcase the power of FSCs with examples in $1d$ and $2d$. In $1d$, we show the FSC of the average cluster state with $\mathbb{Z}_2\times \mathbb{Z}_2^A$ symmetry (``A" denotes average symmetries throughout this paper) is long-range ordered. In $2d$, we demonstrate several examples including bosonic ASPTs with the so-called $0d$-decoration and $1d$-decoration and one example of fermionic ASPT with $1d$-decoration. We uncover an intriguing connection between FSCs and so-called watermelon correlators in $O(n)$ loop models with quantum corrections coming from the decorated domain wall structure of the ASPT states, granting us exact scaling exponents for FSCs. After these theoretical exploration, in Sec. III, we discuss the experimental proposal to measure strange correlators with classical shadow tomography. 

\subsection{A short review for ASPT states} 

We consider general mixed states which can be either the result of quantum decoherence on a pure state or the ensemble of ground states of disordered Hamiltonians. Mixed states can host two distinct types of symmetries. The exact symmetry $K$ is a symmetry for each individual quantum trajectory or disordered Hamiltonian, while the average symmetry $A$ is only a statistical symmetry of the ensemble. Mathematically, an exact symmetry acts on the density matrix as $U_k\rho=e^{i\theta_k}\rho$, while an average symmetry action is $U_a\rho U_a^\dagger=\rho$. 

Physically, an ASPT state should be a short-range correlated state that preserves these symmetries without spontaneous breaking. More precisely, we require the density matrix $\rho$ to be a gapped, symmetrically invertible mixed state -- a concept that is detailed in Refs. \cite{lessa2024mixed, ma2025topological}\footnote{The notion of a gapped mixed state is defined through its entanglement properties. More precisely, a mixed state is considered gapped if both the mutual information and the conditional mutual information between two subsystems, separated by a buffer region, decay exponentially with the size of their separation \cite{lessa2024mixed, sang2024stability}. For a mixed state $\rho$ to be symmetrically invertible, this requires that one can always find another mixed state $\tilde{\rho}$, such that $\rho\otimes\tilde{\rho}$ is two-way connected to a trivial product state by symmetric finite-depth local quantum channels. The gapped symmetrically invertible condition serves as the mixed-state analog of the symmetric short-range entanglement condition for pure-state SPTs. }. Roughly speaking, this condition ensures the absence of any form of symmetry breaking, including ordinary spontaneous symmetry breaking and strong-to-weak symmetry breaking, in the density matrix. To rule out conventional spontaneous symmetry breaking states, we impose that the correlation functions of charged operators (with respect to either the strong or weak symmetry) decay exponentially with distance.
\begin{align}
\label{SSB}
\mathrm{Tr}(\rho O^\dagger(r)O(r'))\sim e^{-|r-r'|/\xi},
\end{align}
Additionally, we must exclude another type of symmetry breaking, known as strong-to-weak symmetry breaking \cite{Lee_2023, ma2025topological, lessa2024strong}. This novel form of symmetry breaking cannot be detected through ordinary correlation functions; instead, it is identified using the following fidelity correlators of 
$K$-charged operators. 
\begin{align}
\label{swssb}
F\left(\rho,O_i^\dag O_j \rho O_j^\dag O_i\right)\sim e^{-|i-j|/\xi},
\end{align}
where $F(\rho,\sigma)=\mathrm{Tr}\sqrt{\sqrt{\rho}\sigma\sqrt{\rho}}$ is the fidelity between density matrices $\rho$ and $\sigma$. Exponential decay of the fidelity correlation function indicates no strong-to-weak symmetry breaking in the mixed state. The exponential decay of both correlation functions in Eqs. \eqref{SSB} and \eqref{swssb} indicates that the density matrix remains symmetric, a necessary condition for the state to be an ASPT.

Nontrivial ASPTs refer to gapped, symmetrically invertible mixed states that cannot be ``smoothly connected"\footnote{In the disordered systems, this corresponds to smoothly varying the ensemble of disordered Hamiltonians. In the case of an open quantum system, it refers to the bidirectional connectivity of the states under quantum channels.} to trivial product states without breaking the exact and average symmetries. While the classification of ASPT states is not the focus of this paper (see Ref. \cite{MaWangASPT, ma2025topological} for more details), a particularly useful physical framework -- the decorated domain wall picture \cite{XieChen_2014} -- effectively captures the essential features of ASPT states. The fixed-point ASPT density matrix can be interpreted as a classical ensemble of symmetry defect configurations associated with the average symmetry $A$. Each $A$ symmetry defect is then decorated with a quantum mechanical state, specifically an SPT state protected by the system's exact symmetry $K$. Generally, if the symmetry defect of average symmetry $A$ is $n$-dimensional, we refer this to as $n$-$d$-decoration pattern. The nontrivial SPT decoration preserves the quantum mechanical nature of the state and prevents it from being smoothly connected to a trivial state without breaking the $K$ symmetry. Moreover, the boundaries of these ASPT states often host symmetry-protected long-range entanglement as a consequence of the nontrivial decorations \cite{MaWangASPT}.

A classification of ASPT states for general onsite symmetries, based on the decorated domain wall framework, has been proposed in Ref. \cite{ma2025topological}. This classification provides crucial insights into the stability of SPTs in mixed states. Specifically, it indicates that many pure-state SPTs remain robust against average symmetric disorder or decoherence, meaning that their topological properties persist even in the presence of disorder or decoherence. More unexpectedly, the classification also identifies a new class of ASPT phases that do not have a corresponding pure-state SPT counterpart. These states, termed intrinsic ASPTs, exhibit topological properties that emerge only in the presence of disorder or decoherence. These findings suggest that ASPTs encompass a broader and richer landscape of phases than previously recognized in pure-state settings. The generalized decorated domain wall framework provides a systematic approach to understanding both the ASPTs that descend from pure-state SPTs and intrinsic ASPTs that exist purely in the mixed states. To detect and characterize these phases, both numerically and potentially in experiments, we introduce the FSC (defined below), which serves as a powerful diagnostic tool for identifying nontrivial ASPTs from both classes. 

\section{Fidelity Strange Correlator} 

The fidelity strange correlator for a mixed state $\rho$ is defined as
\begin{align}
C(r,r')=\frac{F\left(\rho,\phi(r)\phi(r')\rho_0\phi(r)\phi(r')\right)}{F(\rho,\rho_0)},
\label{fidelity strange correlator}
\end{align}
where $\rho_0$ is a reference trivial state preserving the exact and average symmetry. 
It is easy to show that if $\rho$ is a pure-state SPT, the FSC gives exactly the original strange correlator for pure-state SPTs as defined in Eq. \eqref{strange pure}. For instance, in the case of pure states, the fidelity between two states reduces to the wavefunction overlap, which corresponds exactly to the denominator of the strange correlator. Therefore, this form of fidelity strange correlator is a natural generalization of the strange correlator to the mixed-state settings. Although the representation of a density matrix is basis-dependent, we emphasis the definition of the FSC is basis-independent, which is a desirable property for a universal diagnostic tool. 

Nevertheless, to gain an intuitive understanding for the FSC of ASPTs, it is convenient to use a average symmetry breaking basis as the fixed-point density matrix is diagonal in this basis: $\rho=\sum_\D p_\D\ket{\Psi_\D}\bra{\Psi_\D}$. Each $\ket{\Psi_\D}$ is a $K$ symmetric but $A$-breaking wavefunction and the sum over $\D$ takes into account of all possible $A$ symmetry breaking patterns. Physically, each $|\Psi_\D\rangle$ can be regarded as a particular quantum trajectory of decoherence processes or the ground state of a specific symmetry breaking disordered Hamiltonian. For a nontrivial ASPT, each $|\Psi_\D\rangle$ has nontrivial $K$-SPT decorations on the symmetry defects of $A$. For the choice of reference state, we can also utilize the A-symmetry breaking basis: $\rho_0=\sum_\D p_\D\ket{\Phi_\D}\bra{\Phi_\D}$ where, for simplicity, we assume the same probability distribution. Each $|\Phi_\D\rangle$ is a trivial product state preserving the $K$ symmetry (i.e., without decoration). Now the FSC can be reformulated in the following form,
\begin{align}
\nonumber
C(r,r')=&\frac{\sum_\D p_\D|\bra{\Psi_\D}\phi(r)\phi(r')\ket{\Phi_\D}|}{\sum_\D p_\D|\bra{\Psi_\D}\Phi_\D\rangle|}\\
\label{eq:avg-D}
=&\frac{\sum_\D\tilde p_\D |\langle \phi(r)\phi(r')\rangle_S|}{\sum_\D\tilde p_\D}.
\end{align}
In fact, it can be viewed as a statistical average of the absolute values of pure-state strange correlator
\begin{align}
\langle \phi(r)\phi(r')\rangle_S=\frac{\langle\Psi_\D|\phi(r)\phi(r')|\Phi_\D\rangle}{\langle\Psi_D|\Phi_D\rangle},
\end{align}
over the domain wall configurations $\D$, where the classical weight $\tilde p_\D = p_\D|\langle\Psi_\D|\Phi_\D\rangle|$ is a combination of probability distribution $p_\D$ and $|\langle\Psi_\D|\Phi_\D\rangle|$ given by the overlap between the reference state and ASPT wave functions. Physically, the quantum nature of ASPT states originates from the decorated SPTs on domain walls. As a result, wavefunction overlaps play a crucial role in determining the long-wavelength behavior of the strange correlator. We therefore refer to these contributions as \textit{quantum corrections}. As expressed in Eq. \ref{eq:avg-D}, the strange correlator can be interpreted as a specific correlation function in a statistical mechanics model, determined by both the probability distributions and quantum corrections.

In the following, we will show with examples that FSCs have long-ranged or power-law behavior at long distances if $\rho$ describes a nontrivial ASPT.


\subsection{1d example} 

We will start with a simple $1d$ example -- the $1d$ averaged cluster state protected by $\mathbb{Z}_2\times\mathbb{Z}_2^A$ symmetry. This ASPT can be obtained by decoherence on a pure state SPT with $\mathbb{Z}_2\times\mathbb{Z}_2$ symmetry. 
The pure state SPT is stablized by the cluster Hamiltonian $H=-\sum_{j=1}^{2N}Z_{j-1}X_jZ_{j+1}$, which builds in the decorated domain wall structure. The two $\mathbb{Z}_2$ symmetries are defined on the odd and even sites, respectively, 
\begin{align}
\mathbb{Z}_2^{\mathrm{odd}}=\prod_{j\in\mathrm{odd}}X_j,~~\mathbb{Z}_2^{\mathrm{even}}=\prod_{j\in\mathrm{even}}X_j.
\end{align}
For simplicity, we consider strong measurements in the $Z$-basis on the even sites which breaks $\mathbb{Z}_2^{\mathrm{even}}$ down to average. In this case, we can write down the explicit density matrix states, 
\begin{align}
\rho=\sum_\D\frac{1}{2^N}|\Psi_\D\rangle\langle\Psi_\D|,
\end{align}
where 
\begin{align}
|\Psi_\mathcal{D}\rangle=\bigotimes_{j=1}^N|Z_{2j}=\sigma_{2j}^{\D}\rangle\otimes|X_{2j+1}=\sigma_{2j}^{\D}\sigma_{2j+2}^{\D}\rangle,
\end{align}
which follows the decorated domain wall picture. Taking the reference state with $|\Phi_\mathcal{D}\rangle=\bigotimes_{j=1}^N|Z_{2j}=\sigma_{2j}^{\D}\rangle\otimes|X_{2j+1}=1\rangle$, where no charge is attached to the domain wall, we explicitly find the FSC of $Z_i$ operators on even sites long-range correlated\footnote{In the supplementary, we show the FSC of $Z$ operators on odd sites is also long-range and the FSCs are related to the string order parameters of the SPT order.}, namely $C_{ZZ}(i,j)=1$. In the supplementary\cite{supplementary}, we show that the long-ranged behavior goes beyond the strong-measurement limit as long as the $A$ symmetry is preserved on average. 

\subsection{2d examples} 

\begin{figure}
\includegraphics[width=0.4\textwidth]{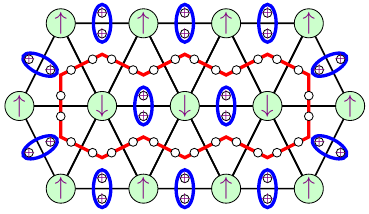}
\caption{Decorated domain wall states for bosonic ASPT with $\mathbb{Z}_2\times \mathbb{Z}_2\times \mathbb{Z}_2^A$ symmetry and fermionic ASPT with $\mathbb{Z}_2^f\times \mathbb{Z}_2^A$ symmetry. Ising spins associated with $\mathbb{Z}_2^A$ symmetry are on the triangular lattice sites (green). For the boson case, each link hosts two spin-1/2's. A 1$d$ cluster state (red) is attached to each Ising domain wall, while in a blue circle, the spins are polarized to $|+\rangle$ states. For fermionic ASPT, the link spins are replaced by Majorana fermions, and the decorated states are replaced by $1d$ Kitaev chains.}
\label{DWdecoration}
\end{figure}

We consider a $2d$ bosonic ASPT  with $\mathbb{Z}_2\times\mathbb{Z}_2 \times\mathbb{Z}_2^A$ symmetry. The ASPT is constructed by decorating the domain wall of the average $\mathbb{Z}_2^A$ symmetry with a $1d$ cluster state protected by the $\mathbb{Z}_2\times\mathbb{Z}_2$ exact symmetry as pictorially shown in Fig. \ref{DWdecoration}. This ASPT exists for both decohered and disordered systems. 

We first consider the FSC of the $Z$ operators of spins on the links which are charged under the exact symmetry $K$, namely 
\begin{equation}
    C_{K}(r,r')=\frac{F\left(\rho,Z_rZ_{r'}\rho_0Z_rZ_{r'}\right)}{F(\rho,\rho_0)}
    \label{2dZ2Z2Z2SC}
\end{equation}
The subscript $K$ denotes the FSC are for $K$-charged operators.
Using Eq.~\eqref{eq:avg-D}, the FSC can be expressed as a statistical average over configurations of the Ising spins, labeled by $\D$.
Without loss of generality, we assume the probability $p_\mathcal{D}$ the form of the Boltzmann weight of a classical Ising model, or equivalently $p_{\mathcal{D}}=2\prod_{l\in\mathcal{D}}x^{L(l)}$, where $x=e^{-2\beta}$ \footnote{The choice of $p_\D$ is of course not unique. However, as long as the Ising degrees of freedom is in the paramagnetic phase, one should be able to write a similar $p_\D$ with a renormalized loop tension.}. Here, $l$ labels a domain wall in the Ising configurations and $L(l)$ its length. For the ensemble to be in a $\mathbb{Z}_2^A$ symmetric phase -- where the two-point correlation function of $\mathbb{Z}_2^A$ charges decays exponentially with distance -- we need $x>x_c^{\mathrm{Ising}}=1/\sqrt{3}$ for a triangular lattice. 
Furthermore, the weight of each configuration is ``quantum corrected''
by the wavefunction overlap $|\langle\Psi_\mathcal{D}|\Phi_\mathcal{D}\rangle|$, which can be decomposed into the product of overlap between a cluster state and a trivial state on each domain wall,
\begin{equation}
\langle\Psi_\mathcal{D}|\Phi_\mathcal{D}\rangle=\prod_{l\in\mathcal{D}} \langle \psi_{\mathrm{cluster}}|\psi_0\rangle (l)=\prod_{l\in\mathcal{D}} 2\times 2^{-L(l)}.
\label{overlap}
\end{equation}
Here the overlap is calculated with the fixed point wavefunction of the cluster state and trivial state. Each overlap factor decays with the length of the domain wall. The decay rate generally is not universal, while the factor of 2 in front of the exponential decay is \textit{universal} -- it equals to the degeneracy of the boundary modes of the decorated $1d$ SPT. We show in the supplementary\cite{supplementary} that this factor stays the same for SPT wavefunction away from the fixed point using both field theory and matrix product representations for $1d$ SPT. This factor is crucial for the behavior of FSCs. Eventually, the ``quantum corrected'' weight can be written as 
\begin{equation}
\tilde p_\mathcal{D}=2\tilde{x}^{L(\mathcal{D})}2^{n(\mathcal{D})},
\label{partition}
\end{equation}
where $L(\mathcal{D})$ is the total length of domain wall, $n(\mathcal{D})$ is the number of domain walls in configuration $\mathcal{D}$, and $\tilde{x}=x/2$ is the renormalized loop tension. Eq. (\ref{partition}) resembles the partition function of an $O(n)$ loop model with loop fugacity $n=2$.  We emphasize that both the loop fugacity and tension have received nontrivial quantum corrections from the decorated domain wall states. 

For the numerator, the crucial observation is that the strange correlator $C_K^\mathcal{D}(r, r')$ is non-zero only if the two measured spins reside on the same domain wall, and the non-zero value is precisely the strange correlator of the $1d$ cluster state.
Therefore, the FSC, in the end, can be written as
\begin{equation}
C_{K}(r,r')=\frac{\sum_{\D'}\langle Z_rZ_{r'}\rangle_S\tilde{x}^{L(\mathcal{D}')}2^{n(\mathcal{D}')}}{\sum_{\D}\tilde{x}^{L(\mathcal{D})}2^{n(\mathcal{D})}},
\label{2DSCwith1deco}
\end{equation}
conditioned on that in every configurations of $\D'$ there must be a domain wall connecting $r$ and $r'$. The factor $\langle Z_rZ_{r'}\rangle_S$ is a non-zero constant as the strange correlator of the $1d$ cluster state. Therefore, the $K$-FSC measures the probability that $r$ and $r'$ are connected by a single loop in the O(2) loop model. This quantity maps exactly to the 2-leg watermelon correlator in the $O(2)$ loop model\cite{Duplantier_1989}. For the loop tension $\tilde{x}>x_c^{n=2}=1/\sqrt{2}$, the loop model will be in the dense loop phases, which indicates the FSC has a power-law behavior:
\begin{equation}
C_{K}(r,r')\sim{|r-r'|^{-2\Delta_2}},
\label{2leg}
\end{equation}
where $\Delta_2$ is known as the 2-leg exponent whose value in the $O(2)$ loop model is $\Delta_2=1/2$.

\begin{figure}
    \centering
    \includegraphics[width=0.45\textwidth]{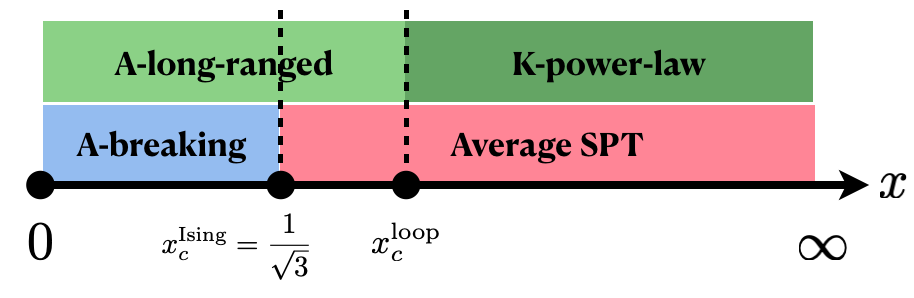}
    \caption{Phase diagram from the FSCs. The ASPT is well defined in $x>x_c^{\text{Ising}}$. For $x<x_c^{\text{Ising}}$, the $A$ symmetry is spontaneously broken.} 
    \label{fig:PD}
\end{figure}

A careful reader may notice that, if we start from the bare loop tension $x\in[1/\sqrt{3},\sqrt{2}]$, then the renormalized loop tension $\tilde{x}$ is not in the dense loop phase. In fact, loop tension in this regime will flow to 0 in the infrared and the system prefers to have no loops, also known as the dilute loop phase. Therefore, the FSC defined in Eq. (\ref{2dZ2Z2Z2SC}) decays exponentially with distance. However, this does not indicate that the FSC fails to detect the ASPT order because we also need to include the FSC associated with operators with nontrivial $\Z_2^A$-charge, namely
\begin{equation}
C_A(r,r')=\frac{F(\rho,\sigma_r\sigma_{r'}\rho_0\sigma_r\sigma_{r'})}{F(\rho,\rho_0)},
\label{GSC}
\end{equation}
where $\sigma$'s are the Ising spins on the sites of the triangular lattice. The FSC in fact measures the probability of two points sitting in the same domain. In the dilute loop phase where loops are suppressed, this correlator is actually long-ranged,
\begin{equation}
    C_A(r,r')\sim \text{const},\ \  \text{for}\ \  \tilde{x}<1/\sqrt{2}.
\end{equation}
Therefore, FSCs are either long-ranged or power-law in the whole regime wherever the ASPT is well-defined, as shown in the phase diagram in Fig. \ref{fig:PD} \footnote{We note that the ASPT state is only well-defined in the regime $x>x_c^{\rm Ising}$. For $x<x_c^{\rm Ising}$, the $A$ symmetry is spontaneously broken. }.

We also construct an example of fermionic ASPT in $2d$ with $1d$-decoration. The example is the averaged version of $2d$ fermionic SPT with unitary $\mathbb{Z}_2$ symmetry \cite{KITAEVanyon}. The decorated domain wall wavefunction is similar to the boson case shown in Fig. \ref{DWdecoration}. On the links of the triangular lattice, we put 2 Majorana modes, labeled $\chi_A$ and $\chi_B$, forming a complex fermion $c=(\chi_A+i\chi_B)/2$. For Majorana modes on an Ising domain wall, they form a $1d$ topological superconductor \cite{1Dfermion,2DTSC, DWfermion1, DWfermion2}. Otherwise, they pair up locally to even parity states.

Consider the FSC of the $c$ fermions,
\begin{equation}
C_K(r,r')=\frac{F(\rho,c(r)c(r')\rho_0 c^\dag(r')c^\dag(r))}{F(\rho,\rho_0)}. 
\end{equation}
The essential difference from the previous case is the quantum correction of the loop fugacity from the SPT-trivial wavefunction overlap on the domain walls. Indeed, we can show for the fixed point wavefunction, the overlap has the form $\langle\Psi_\mathcal{D}|\Phi_\mathcal{D}\rangle=\prod_{l\in\mathcal{D}} \langle \psi_{\mathrm{Majorana}}|\psi_0\rangle (l)=\prod_{l\in\mathcal{D}} \sqrt{2}\times \sqrt{2}^{-l}$ \cite{supplementary}. This means that the loop model now has loop fugacity $n=\sqrt{2}$. For the numerator, similar to the previous case, the only configurations that are non-zero need to have one domain wall going through the two fermion positions. Therefore, the FSC reduces to 
\begin{equation}
C_K(r,r')=\frac{\sum_{\D'}\langle c_rc_{r'}\rangle_S\tilde{x}^{L(\mathcal{D}')}\sqrt{2}^{n(\mathcal{D}')}}{\sum_{\D}\tilde{x}^{L(\mathcal{D})}\sqrt{2}^{n(\mathcal{D})}},
\label{2DSCwith1deco}
\end{equation}
where $\langle c_rc_{r'}\rangle_S$=const is the strange correlator of the $1d$ Kitaev chain. The FSC maps to the 2-leg watermelon correlator in the $O(n=\sqrt{2})$ model. For $\tilde{x}>x_c^{n=\sqrt{2}}\cong 0.601$, the loop model falls into the dense loop fixed point, where the FSC has a similar power-law behavior as in Eq. (\ref{2leg}) with an exponent $\Delta_2=1/3$. For $\tilde{x}=x_c^{n=\sqrt{2}}$, the loop model is at the dilute fixed point, where the exponent becomes $\Delta_2=3/5$. For $\tilde{x}<x_c^{n=\sqrt{2}}$, the $K$-FSC becomes short-ranged. In this regime, we consider the $A$-FSC defined as Eq. (\ref{GSC}). Since the loop model is in a dilute loop phase, the $A$-FSC is long-range ordered. 

\begin{figure}
\includegraphics[width=0.4\textwidth]{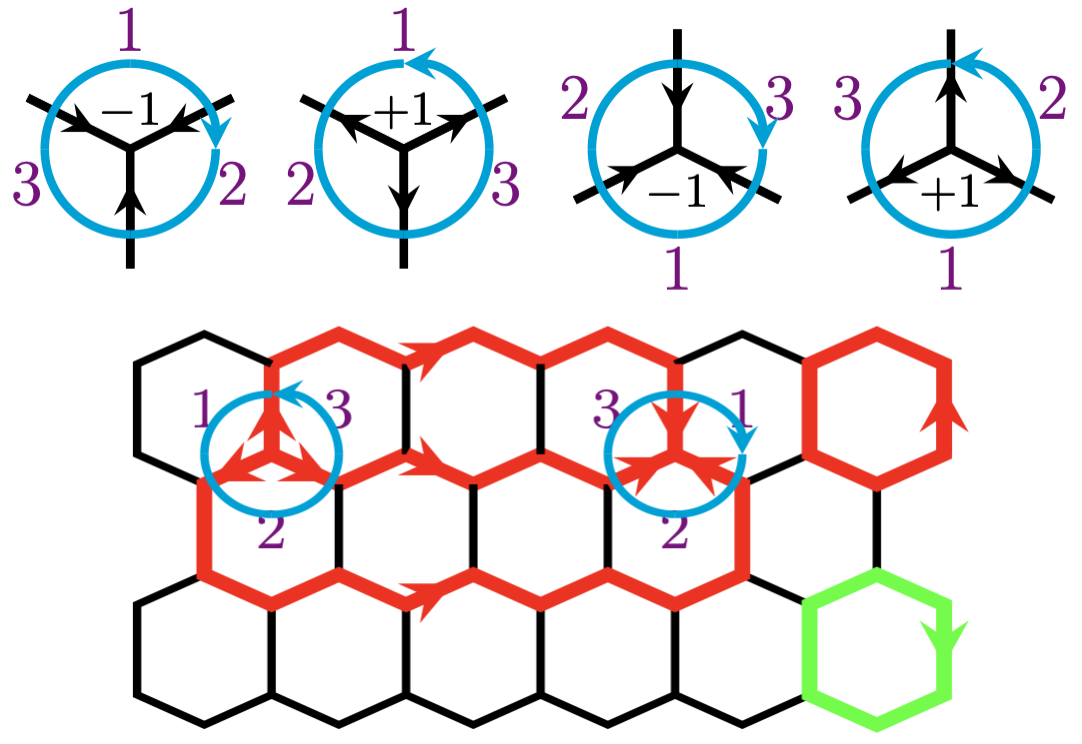}
\caption{Top: the charge decoration rule for $\mathbb{Z}_3\times \mathbb{Z}_3^A$ symmetry. A $\pm2\pi$ vortex of $\mathbb{Z}_3^A$ is decorated with $\pm1$ charge of $\mathbb{Z}_3$. Bottom: The allowed configuration in the numerator of the FSC in Eq. (\ref{SCZ3}). }
\label{vortexdeco}
\end{figure}

After considering $2d$ cases with $1d$-decoration, we give an example of $2d$ decohered bosonic ASPT with $0d$-decoration with $\mathbb{Z}_3\times \mathbb{Z}_3^A$ symmetry. This ASPT has a decoration structure in that a nontrivial $\mathbb{Z}_3$ is attached on each vortex of the $\mathbb{Z}_3^A$ order parameter as shown in Fig. \ref{vortexdeco}. The ensemble of states consistent with the decoration rule defines a $\mathbb{Z}_3\times\mathbb{Z}_3^A$ ASPT. 

We consider the strange correlation of creation and annihilation operators of $\mathbb{Z}_3$ charges, namely,
\begin{equation}
C_K(r,r')=\frac{F(\rho,a^\dag(r)a(r')\rho_0 a^\dag(r')a(r))}{F(\rho,\rho_0)}.
\label{SCZ3}
\end{equation}
We specify that the reference state has 0 charges decorated on the $\mathbb{Z}_3^A$ vortices. $\D$ again represents all different configurations of $\mathbb{Z}_3^A$ order parameter. In particular, there will be domain wall and vortex configurations. However, if $\D$ has vortices, then the overlap function between the trivial and SPT wavefunction in the denominator will be identically zero due to the nontrivial $\mathbb{Z}_3$ global charge decorated on the vortex. Therefore, all the vortex configurations are killed in the summation and we end up again with a loop model. Note that there are two flavors of loops since there are two different kinds of domain walls in a $\mathbb{Z}_3$ model. Here, without loss of generality, we assume the probability distribution is given by a thermal weight of a $\mathbb{Z}_3$ clock model.
As a result, the loop tensions for the two kinds of domain walls are the same. The denominator can be written as $
F(\rho,\rho_0)=3\sum_{\mathcal{D}'}x^{L(\mathcal{D'})}2^{n(\mathcal{D'})}$, where $\D'$ only contain loop configurations and the overall factor of $3$ comes from the $\mathbb{Z}_3$ symmetry. The denominator again maps to the partition function of an $O(2)$ loop model. 

By the same logic, the summation in the numerator of Eq. (\ref{SCZ3}) should also contain only loop configurations except there should be a vortex and an anti-vortex right at position $r$ and $r'$ respectively due to the charge creation and annihilation operators in the definition of Eq. (\ref{SCZ3}). We demonstrate examples of allowed configuration in Fig. \ref{vortexdeco}. In this case, the FSC maps to the 3-leg watermelon correlation of the $O(2)$ model, which gives 
\begin{equation}
    C_K(r,r')\sim {|r-r'|^{-2\Delta_3}},
\end{equation}
where $\Delta_3=9/8$ for $x>x_c^{{n=2}}=1/\sqrt{2}$. Again, if the loop tension is $x<1/\sqrt{2}$, then the loop model is in the dilute loop phase and we can measure the analogous $A$-charge FSC to find long-range order. 

\section{Experimental measurement of fidelity strange correlators}

{
\color{black}
The aim of developing the FSC is to detect a nontrivial ASPT states directly from a bulk density matrix. Therefore, one central question is how we can measure this FSC experimentally on a quantum device. In this section, we provide a quantum metrological way of systematically estimating the FSC by classical shadow tomography in experiments\cite{Huang_2020, Rath_2021, Hu_2023}. 

On the one hand, the idealist way to experimentally probe the FSC \eqref{fidelity strange correlator} requires precise measurements of the fidelity. Because of the highly nonlinear nature of the fidelity (and any other quantum information-theoretic quantities), a precise measurement of the fidelity generally requires a quantum state tomography (QST) to fully reproduce all information of the density matrix. Nevertheless, the QST is exponentially hard with respect to the system size, therefore, the rigorous experimental probe of FSC is almost impossible. 

On the other hand, although a precise measurement is not possible, a reliable estimation is still desirable. Here, we propose a reliable estimation of FSC and the corresponding experimental measurements in quantum devices. 

In Ref. \cite{miszczak2008sub}, it was proven that the fidelity of two density matrices $\rho$ and $\sigma$ is lower bounded by the \textit{sub-fidelity}, namely
\begin{align}
F(\rho,\sigma)\geq\mathrm{Tr}(\rho\sigma)+\sqrt{2}\sqrt{\left[\mathrm{Tr}(\rho\sigma)\right]^2-\mathrm{Tr}\left[(\rho\sigma)^2\right]}.
\end{align}
Inspired by the above lower bound, in Ref. \cite{zhang2024fluctuation}, a series expansion of the fidelity has been proposed in the following form,
\begin{align}
F=F_1+\sqrt{2F_2+\sqrt{8F_3+\sqrt{\cdots\sqrt{2^{\frac{n(n-1)}{2}F_n+\cdots}}}}},
\label{Eq: series}
\end{align}
where $F_1=\mathrm{Tr}(\rho\sigma)$, and
\begin{align}
F_{n+1}=F_n^2-2^{\frac{n(n-1)}{2}}\left(\left[\mathrm{Tr}\left(\rho\sigma\right)^{2^{n-1}}\right]-\mathrm{Tr}\left[(\rho\sigma)^{2^n}\right]\right).
\label{Eq: Fn}
\end{align}
We note that each term $F_n$ for arbitrary $n\in\mathbb{Z}^+$ is a polynomial of the density matrices, hence each term is experimentally measurable by the method of classical shadow tomography \cite{Huang_2020, Rath_2021, Hu_2023}. 

The idea of classical shadow tomography is to efficiently characterize or “learn” properties of an unknown quantum state using relatively few measurements. These “shadows” are compact classical descriptions obtained from randomized measurements, which can then be used to estimate many different observables’ expectation values. For simplicity, we will consider qubit systems. 
we can estimate the density matrix $\sigma=\phi(r)\phi(r')\rho_0\phi(r)\phi(r')$ by performing the randomized measurements on the density matrices.  Specifically, we apply a random unitary $U$ to the state $\sigma$ and measure in the computational basis. The outcome is a bit string state $\ket{\boldsymbol{b}}$, $\boldsymbol{b}\in \{0,1\}^{\otimes N}$. The post-measurement density matrix is $\ket{\boldsymbol{b}}\bra{\boldsymbol{b}}$, with probability $p_{U,\boldsymbol{b}}=Tr(U \sigma U^\dagger \ket{\boldsymbol{b}}\bra{\boldsymbol{b}})$ given by the Born's rule. From each measurement outcome, we construct a classical shadow (an operator) denoted by $\tau=U^\dagger\ket{\boldsymbol{b}}\bra{\boldsymbol{b}} U$ such that
\begin{align}
\mathop{\mathbb{E}}\limits_{\tau\sim p(\tau|\sigma)}\tau=\mathcal{M}[\sigma],
\label{Eq: CST}
\end{align}
where $\mathbb{E}$ represents the expectation value of different randomized measurement outcomes, $\mathcal{M}$ represents the measurement channel which is a linear map. We can estimate the original density matrix by reversing this map and obtain 
\begin{align}
\sigma\simeq\mathop{\mathbb{E}}\limits_{\tau\sim p(\tau|\sigma)}\mathcal{M}^{-1}[\tau].
\end{align}
One can similarly reconstruct the other density matrix by random measurements. As a result, according to the rule of classical shadow tomography, we can effectively probe the leading order of the numerator of Eq. \eqref{fidelity strange correlator}, namely
\begin{align}
\mathrm{Tr} & \left(\rho\phi(r)\phi(r')\rho_0\phi(r)\phi(r')\right)\nonumber\\
&=\mathop{\mathbb{E}}\limits_{\substack{\tau\sim p(\tau|\rho) \\ \tau'\sim p(\tau'|{\sigma})}}\mathrm{Tr}\left( \mathcal{M}^{-1}[\tau] \mathcal{M}^{-1}[\tau']\right),
\end{align}
and the leading order of the denominator of Eq. \eqref{fidelity strange correlator} is
\begin{align}
\mathrm{Tr}(\rho\rho_0)=\mathop{\mathbb{E}}\limits_{\substack{\tau\sim p(\tau|\rho) \\ \tau'\sim p(\tau'|{\rho_0})}}\mathrm{Tr}\left( \mathcal{M}^{-1}[\tau] \mathcal{M}^{-1}[\tau']\right).
\end{align}
Therefore, the leading order approximation of the FSC can be experimentally probed by classical shadow tomography, namely
\begin{align}
C^{(2)}(r,r')&=\frac{\mathrm{Tr}\left(\rho\phi(r)\phi(r')\rho_0\phi(r)\phi(r')\right)}{\mathrm{Tr}(\rho\rho_0)}\nonumber\\
&=\frac{\mathop{\mathbb{E}}\limits_{\substack{\tau\sim p(\tau|\rho) \\ \tau'\sim p(\tau'|{\sigma})}}\mathrm{Tr}\left( \mathcal{M}^{-1}[\tau] \mathcal{M}^{-1}[\tau']\right)}{\mathop{\mathbb{E}}\limits_{\substack{\tilde\tau\sim p(\tilde\tau|\rho) \\ \tilde\tau'\sim p(\tilde\tau'|{\rho_0})}}\mathrm{Tr}\left( \mathcal{M}^{-1}[\tilde\tau] \mathcal{M}^{-1}[\tilde\tau']\right)}.
\label{Eq: Type-2}
\end{align}
We note that the leading order approximation of the FSC by the proposed series expansion \eqref{Eq: series} is exactly the type-II strange correlator proposed in Ref. \cite{DSPT}. 

However, as we mentioned before, each term $F_n$ [cf. Eq. \eqref{Eq: Fn}] is a polynomial of the density matrix $\rho$ and $\rho_0$, which is also experimentally measurable by classical shadow tomography. Therefore, the proposed series expansion of the fidelity \eqref{Eq: series} can systematically enhance the estimation accuracy of the FSC. 

}

\section{Conclusion and Discussion} 

In this work, we devise a fidelity strange correlator that can detect nontrivial ASPT phases in disordered or open quantum systems. In $2d$, the FSCs considered here map to correlation functions in certain $2d$ loop models. Remarkably, the decorated domain wall states give nontrivial quantum corrections to the loop tension and loop fugacity -- similar observation was made in strange correlators of pure-state SPTs in Ref. \cite{Scaffidi_2016} with exact solvable models. For intrinsic ASPTs, the structure of the FSCs and the mapping to statistical models are expected to be the same since the decorated domain wall construction is still valid for intrinsic ASPTs. We also provide a systematic way of estimating the FSC experimentally, through classical shadow tomography. 

The generalization of FSCs to higher dimensional ASPTs is an exciting future direction. The map from FSCs to statistical models is not limited to $2d$. For $3d$ ASPT with $1d$-decoration, the resulting statistical model will be $3d$ loop models for which some analytical results are also known \cite{Cardy_2005, Nahum_2011, Nahum_2015}. For $2d$-decoration, for example, a $3d$ ASPT phase with $K=\mathbb{Z}_2$ and $A=\mathbb{Z}_2^A$ from decorating a 2D Levin-Gu state on the codimension-1 $\mathbb{Z}_2^A$ domain wall, the FSCs should be mapped to correlation functions in $3d$ membrane models. However, the effect of the decorated 2d state on the membrane model is unclear.


\begin{acknowledgments}
\textit{Acknowledgements} -- We thank Yimu Bao, Meng Cheng, Adrian Po, Chong Wang, Shuo Yang, Yi-Zhuang You, Carolyn Zhang, and Jing-Yu Zhao for stimulating discussions, and especially Adam Nahum for discussing the loop model and their critical behaviors. J.H.Z. and Z.B. are supported by startup funds from the Pennsylvania State University. Z.B. also acknowledges support from NSF under award number DMR-2339319. Y.Q. acknowledges support from the National Key R\&D Program of China (Grant No. 2022YFA1403402) and from the National Natural Science Foundation of China (Grant No. 12174068). 

\textit{Note Added}: While finishing up this work, we became aware of an independent work \cite{DSPT} which considers other generalizations of strange correlators for SPT states under decoherence.
\end{acknowledgments}

\appendix

\section{fidelity strange correlator of 1$d$ average cluster state}
In this section, we explicitly calculate the fidelity strange correlator of the 1$d$ average cluster state. The stabilizer for a pure 1$d$ cluster state is
\begin{align}
H=-\sum\limits_{j=1}^{2N}Z_{j-1}X_jZ_{j+1},
\end{align}
where $Z$ and $X$ are Pauli matrices that are anti-commute. There is a $\Z_2\times\Z_2$ symmetry that is defined on the odd and even sites, respectively, with the generators as
\begin{align}
\Z_2^{\rm odd}=\prod\limits_{j\in \mathrm{odd}}X_j,~\Z_2^{\rm even}=\prod\limits_{j\in \mathrm{even}}X_j.
\end{align}
Then we perform strong $Z$ measurements on the even sites which can be done by coupling each qubit on an even site with an ancilla qubit through a controlled-$Z$ gate and then tracing out the ancilla. The $\Z_2^{\rm even}$ symmetry is broken to an average symmetry $\Z_2^A$ in the resulting mixed state. The mixed state is an ensemble of states, each of which is given by the form
\begin{align}
|\Psi_{\D}\rangle=\bigotimes\limits_{j=1}^N|Z_{2j}=\sigma_{2j}^{\D}\rangle\otimes|X_{2j+1}=\sigma_{2j}^{\D}\sigma_{2j+2}^{\D}\rangle,
\label{psi}
\end{align}
where $\{\sigma_{2j}\}$ are measurement outcomes. We can see that for each specific measurement outcome, the corresponding wavefunction has an explicitly decorated domain wall structure. The trivial reference wavefunctions are chosen as following
\begin{align}
|\Phi_\D\rangle=\bigotimes\limits_{j=1}^N|Z_{2j}=\sigma_{2j}^\D\rangle\otimes|X_{2j+1}=1\rangle,
\label{phi}
\end{align}
where there is no charge decorated to the domain wall. 

In the strong measurement limit, we have equal probability for each state. Here, we can consider a more general probability distribution, $p_\D\sim e^{-n_{\rm DW}\beta J}$, where $n_{\rm DW}$ is the number of domain walls of even spins, and $J$ can be thought of as the energy cost of creating a domain wall. This is a Boltzmann factor of a 1d classical Ising model at finite temperature. As long as $\beta<\infty$, the Ising model is in a disordered phase, thus, the ensemble preserves the average $\mathbb{Z}_2$ symmetry. With this general probability distribution, we can calculate the fidelity strange correlator of the $Z$ operators on the even sites, 
\begin{align}
C_{ZZ}(i,j)=\frac{F(\rho,Z_iZ_j\rho_0Z_iZ_j)}{F(\rho,\rho_0)},
\label{1d fidelity strange correlator}
\end{align}
Let us first consider the denominator, which can be expressed as the summation of wavefunction overlaps as $F(\rho,\rho_0)=\sum_\D p_\D|\langle\Psi_\D|\Phi_\D\rangle|$ in the decorated domain wall basis. We note that according to the explicit expressions in Eqs. (\ref{psi}) and (\ref{phi}), the wavefunction overlap $\langle\Psi_\D|\Phi_\D\rangle$ is nonzero if and only if $\sigma_{2j}^\D \sigma_{2j+2}^\D=1$, $\forall j$. It turns out that the denominator of the fidelity strange correlator should be 
\begin{align}
F(\rho,\rho_0)=2e^{-\beta E_0},
\label{partitionfunction}
\end{align}
where $E_0$ is the ground state energy of the 1$d$ Ising model of the qubits on even sites.

Then we focus on the numerator of the fidelity strange correlator. We first consider $Z_i$ and $Z_j$ operators on the even sites. The $Z$'s will extract the measurement outcomes $\sigma_i^\D$ and $\sigma_j^\D$ on the sites $i$ and $j$. It turns out that the numerator of the fidelity strange correlator should be
\begin{align}
F(\rho,Z_iZ_j\rho_0Z_iZ_j)=2e^{-\beta E_0}.
\end{align}
Therefore, the fidelity strange correlator $C_{ZZ}(i,j)=1$ for general $\beta<\infty$, which implies that the decohered average 1$d$ cluster state is a nontrivial ASPT state.

We can also derive the nontrivial fidelity strange correlator from the string order parameter. Suppose $i=2i'$ and $j=2j'$, we consider the string order parameter to be defined as
\begin{align}
S_1=Z_{2i'}\left(\prod\limits_{k=i'}^{j'-1}X_{2k+1}\right)Z_{2j'},
\label{string}
\end{align}
where the operator in the bracket is the truncated exact symmetry operator. For any $\beta<\infty$, it is easy to check the expectation value of the above string order parameter is equal to 1, 
\begin{align}
\mathrm{Tr}\left(\rho S_1\right)=1.
\end{align}

This is because that each wavefunction $|\Psi_\D\rangle$ in $\rho$ is invariant by acting the string order parameter. 
Now consider each term in the numerator of the fidelity strange correlator
\begin{align}
\langle\Psi_\D|Z_{2i'}Z_{2j'}|\Phi_\D\rangle=\langle\Psi_\D|Z_{2i'}\left(\prod\limits_{k=i'}^{j'-1}X_{2k+1}\right)Z_{2j'}|\Phi_\D\rangle,
\end{align}
because $|\Phi_\D\rangle$ is invariant under the truncated symmetry operator. Furthermore, each wavefunction $|\Psi_\D\rangle$ is invariant under the string operator. Therefore, we have
\begin{align}
\langle\Psi_\D|Z_{2i'}Z_{2j'}|\Phi_\D\rangle=\langle\Psi_\D|\Phi_\D\rangle.
\end{align}
With this, we can see the fidelity strange correlator equals $1$ for any $\beta<\infty$. 

Now we consider the fidelity strange correlator of $Z_{2i-1}$ and $Z_{2j+1}$, namely $Z$ operators on odd sites. We can show explicitly the fidelity strange correlator is long-range correlated in this case as well. We will make a connection with a fidelity version of the string order parameter. The denominator of the fidelity strange correlator is the same as in Eq. \eqref{partitionfunction}. The numerator of \eqref{1d fidelity strange correlator} can be calculated explicitly: $Z_{2i-1}/Z_{2j+1}$ flips the $X$ eigenvalues at the site-($2i-1$)/($2j+1$), therefore, the wavefunction overlap $\langle\Psi_\D|Z_{2i-1}Z_{2j+1}|\Phi_\D\rangle$ is nonzero if and only if there were exactly two domain walls, $\sigma_{2i-2}\sigma_{2i}=-1$ and $\sigma_{2j}\sigma_{2j+2}=-1$. It turns out that the numerator of \eqref{1d fidelity strange correlator} should be
\begin{align}
F(\rho,Z_{2i-1}Z_{2j+1}\rho_0Z_{2i-1}Z_{2j+1})=2e^{-\beta(E_0+2J)},
\label{exact fidelity strange correlator}
\end{align}
and thus the fidelity strange correlator $C_{ZZ}(2i-1,2j+1)=e^{-2\beta J}$ is finite at any $\beta<\infty$. In particular, at $\beta=0$, the fidelity strange correlator is 1. Therefore, we can determine a nontrivial ASPT density matrix by the fidelity strange correlator of both the average and exact degrees of freedom.

Next we consider the connection with the string order parameter associated with the average symmetry. We know that the usual string order parameter defined by $\mathrm{Tr}\left(\rho S_2\right)$, with 
\begin{align}
S_{2}=Z_{2i-1}\left(\prod\limits_{k=i}^jX_{2k}\right)Z_{2j+1},
\label{alternative string}
\end{align}
decays exponentially with distance\cite{MaWangASPT}. However, we can consider a fidelity version of the string order parameter, defined as
\begin{align}
F\left(\rho,S_2\rho S_2\right). 
\label{Fidelity of rho}
\end{align}
This string order parameter can be shown to be exactly equal to 1 when we consider the fixed-point density matrix. In particular, we can show if the string order parameter is 1, the fidelity strange correlator has to be 1 as well. We can see from the following argument. In the decorated domain wall basis, it is easy to show that the explicit form of Eq. \eqref{Fidelity of rho} is expressed as
\begin{align}
F(\rho,S_2\rho S_2)=\sum_{\D}\sqrt{p_\D p_{\D+2}},
\end{align}
where $\D+2$ and $\D$ are different by two domain walls at $2i$ and $2j$. Furthermore, by utilizing the mean inequalities, 
\begin{align}
\sum_{\D}\sqrt{p_\D p_{\D+2}}\leq\sum_{\D}\frac{p_{\D}+p_{\D+2}}{2}=1,
\end{align}
where the equality is taken if and only if $p_{\D+2}=p_\D$ for $\forall \D$. On the other hand, the fidelity strange correlator $C(2i-1,2j+1)=\sqrt{p_{\D_0+2}/p_{\D_0}}$ where $\D_0$ is the spin configuration with no domain wall. Hence the fidelity strange correlator is equal to 1 when the string order parameter is 1. 

\section{fidelity strange correlators of ASPT and clean SPT\label{Sec. purification}}
In this section, we show that the fidelity strange correlator of ASPT phases is equivalent to that of SPT in the pure state assuming the ASPT phase has a clean limit. 

As mentioned in the introduction, for clean SPT the fidelity strange correlator is defined as
\begin{align}
C(r,r')=\frac{\langle\Omega|\phi(r)\phi(r')|\Psi\rangle}{\langle\Omega|\Psi\rangle},
\label{exact fidelity strange correlator}
\end{align}
where $\phi(r)$ is some local operator at the position $r$, $|\Psi\rangle$ is the wavefunction of the SPT phase, and $|\Omega\rangle$ is the wavefunction of a trivial product state. For ASPT the fidelity strange correlator is defined as
\begin{align}
C(r,r')=\frac{F\left(\rho, \phi(r)\phi(r')\rho_0\phi^\dag(r')\phi^\dag(r)\right)}{F(\rho,\rho_0)}.
\label{SCdef}
\end{align}
where $\rho_0$ is the reference trivial density matrix with the same symmetry class, and $F(\rho,\rho_0)=\mathrm{Tr}\sqrt{\sqrt{\rho}\rho_0\sqrt{\rho}}$ is the fidelity between two density matrices $\rho$ and $\rho_0$. In particular, if both $\rho$ and $\rho_0$ are pure states, the fidelity is collapsed to the modulo of wavefunction overlap. 

Assuming the ASPT has a clean limit, we can always obtain a purified clean SPT wavefunction from the mixed-state density matrix that defines the ASPT state. In particular, in the decorated domain wall basis, we can have 
\begin{equation}
    \rho=\sum_\D p_\D |\Psi_\D\rangle\langle\Psi_\D| \longrightarrow |\Psi\rangle = \sum_{\D}\sqrt{p_\D}e^{i\theta_\D}|\Psi_\D\rangle,
   \label{purified}
\end{equation}
where the $\theta_\D$ is a Berry phase that should be self-consistently determined using the $F$-move of $A$-defects. For an ASPT with a clean limit, this procedure can always be done and the resulting wavefunction is guaranteed to be short-range entangled. However, if the ASPT does not have a clean limit, namely an intrinsic ASPT, one cannot find a consistent assignment of the Berry phases. It is easy to come up with a trivial reference state 
\begin{align}
|\Omega\rangle=\sum_{\D}\sqrt{p_\D}|\Phi_{\D}\rangle,
\label{purified wavefunction}
\end{align}
where $|\Phi_\D\rangle$ are just decorating trivial product states on the $A$ symmetry defects. 

The numerator of the fidelity strange correlator in Eq. \eqref{SCdef} is
\begin{align}
F(\rho,\phi(r)\phi(r')\rho_0\phi^\dag(r')\phi^\dag(r))=\sum\limits_{\D}p_{\D}|\langle\Phi_{\D}|\phi(r)\phi(r')|\Psi_{\D}\rangle|.
\end{align}
On the other hand, the numerator of the fidelity strange correlator of the clean SPT phases (\ref{exact fidelity strange correlator}) from the purified wavefunctions (\ref{purified}) and (\ref{purified wavefunction}) is
\begin{align}
\langle\Omega|\phi(r)\phi(r')|\Psi\rangle=\sum\limits_{\D}p_{\D}\langle\Phi_{\D}|\phi(r)\phi(r')|\Psi_{\D}\rangle e^{i\theta_\D}.
\end{align}
Each term of the numerators of the fidelity strange correlators (\ref{SCdef}) and (\ref{exact fidelity strange correlator}) is equivalent up to a phase factor that is the Berry phase acquired from the $F$-move of $A$-defect. Following similar calculations, each term of the denominators of (\ref{SCdef}) and (\ref{exact fidelity strange correlator}) is also equivalent up to this phase. 

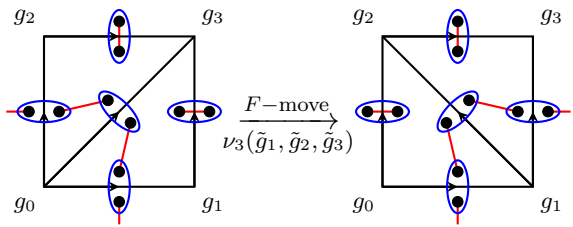
\begin{figure}
\begin{tikzpicture}[scale=1]
\tikzstyle{sergio}=[rectangle,draw=none]
\draw[thick,color=red] (0.2,1) -- (0.85,1.15);
\draw[thick,color=red] (1,0.2) -- (1.15,0.85);
\draw[thick,color=red] (-0.2,1) -- (-0.5,1);
\draw[thick,color=red] (1,-0.2) -- (1,-0.5);
\draw[thick,color=red] (1.8,1) -- (2.2,1);
\draw[thick,color=red] (1,1.8) -- (1,2.2);
\draw[thick] (0,0) -- (0,2) -- (2,2) -- (2,0) -- cycle;
\draw[thick,color=red] (6.3,1) -- (5.65,1.15);
\draw[thick,color=red] (5.5,0.2) -- (5.35,0.85);
\draw[thick,color=red] (5.5,-0.2) -- (5.5,-0.5);
\draw[thick,color=red] (6.7,1) -- (7,1);
\draw[thick,color=red] (4.7,1) -- (4.3,1);
\draw[thick,color=red] (5.5,1.8) -- (5.5,2.2);
\draw[thick] (0,0) -- (2,2);
\draw[thick,->] (0,0) -- (1,0);
\draw[thick,->] (0,0) -- (0,1);
\draw[thick,->] (0,0) -- (1,1);
\draw[thick,->] (0,2) -- (1,2);
\draw[thick,->] (2,0) -- (2,1);
\path (-0.25,-0.25) node [style=sergio] {$g_0$};
\path (2.25,-0.25) node [style=sergio] {$g_1$};
\path (-0.25,2.25) node [style=sergio] {$g_2$};
\path (2.25,2.25) node [style=sergio] {$g_3$};
\filldraw[fill=black, draw=black] (-0.2,1)circle (2pt);
\filldraw[fill=black, draw=black] (0.2,1)circle (2pt);
\filldraw[fill=black, draw=black] (1.8,1)circle (2pt);
\filldraw[fill=black, draw=black] (2.2,1)circle (2pt);
\filldraw[fill=black, draw=black] (1,-0.2)circle (2pt);
\filldraw[fill=black, draw=black] (1,0.2)circle (2pt);
\filldraw[fill=black, draw=black] (1,1.8)circle (2pt);
\filldraw[fill=black, draw=black] (1,2.2)circle (2pt);
\filldraw[fill=black, draw=black] (0.85,1.15)circle (2pt);
\filldraw[fill=black, draw=black] (1.15,0.85)circle (2pt);
\draw[color=blue, thick] (1,0) ellipse [x radius=4pt, y radius=10pt];
\draw[color=blue, thick] (0,1) ellipse [x radius=10pt, y radius=4pt];
\draw[color=blue, thick] (1,2) ellipse [x radius=4pt, y radius=10pt];
\draw[color=blue, thick] (2,1) ellipse [x radius=10pt, y radius=4pt];
\draw[color=blue, thick,rotate around={315:(1,1)}] (1,1) ellipse [x radius=10pt, y radius=4pt];
\path (3.25,1) node [style=sergio] {\large$\xrightarrow{F-\mathrm{move}}$};
\draw[thick] (4.5,0) -- (4.5,2) -- (6.5,2) -- (6.5,0) -- cycle;
\draw[thick] (4.5,2) -- (6.5,0);
\draw[thick,->] (4.5,0) -- (5.5,0);
\draw[thick,->] (4.5,0) -- (4.5,1);
\draw[thick,->] (6.5,0) -- (5.5,1);
\draw[thick,->] (6.5,0) -- (6.5,1);
\draw[thick,->] (4.5,2) -- (5.5,2);
\path (4.25,-0.25) node [style=sergio] {$g_0$};
\path (6.75,-0.25) node [style=sergio] {$g_1$};
\path (4.25,2.25) node [style=sergio] {$g_2$};
\path (6.75,2.25) node [style=sergio] {$g_3$};
\filldraw[fill=black, draw=black] (-0.2+4.5,1)circle (2pt);
\filldraw[fill=black, draw=black] (0.2+4.5,1)circle (2pt);
\filldraw[fill=black, draw=black] (1.8+4.5,1)circle (2pt);
\filldraw[fill=black, draw=black] (2.2+4.5,1)circle (2pt);
\filldraw[fill=black, draw=black] (1+4.5,-0.2)circle (2pt);
\filldraw[fill=black, draw=black] (1+4.5,0.2)circle (2pt);
\filldraw[fill=black, draw=black] (1+4.5,1.8)circle (2pt);
\filldraw[fill=black, draw=black] (1+4.5,2.2)circle (2pt);
\filldraw[fill=black, draw=black] (1.15+4.5,1.15)circle (2pt);
\filldraw[fill=black, draw=black] (0.85+4.5,0.85)circle (2pt);
\draw[color=blue, thick] (1+4.5,0) ellipse [x radius=4pt, y radius=10pt];
\draw[color=blue, thick] (0+4.5,1) ellipse [x radius=10pt, y radius=4pt];
\draw[color=blue, thick] (1+4.5,2) ellipse [x radius=4pt, y radius=10pt];
\draw[color=blue, thick] (2+4.5,1) ellipse [x radius=10pt, y radius=4pt];
\draw[color=blue, thick,rotate around={45:(1+4.5,1)}] (1+4.5,1) ellipse [x radius=10pt, y radius=4pt];
\path (3.25,0.6) node [style=sergio] {$\nu_3(\tilde{g}_1,\tilde{g}_2,\tilde{g}_3)$};
\end{tikzpicture}
\caption{$F$-move of the $A$-defect decorated with $1d$ $K$-SPT decoration. Black solid dots depict the projective representations of $K$ symmetry, and the red links depict their entanglements.}
\label{F-move}
\end{figure}

We demonstrate that if there is no extension between $K$ and $A$, the phase factor $e^{i\theta_\D}$ in the $F$-symbol illustrated in Fig. \ref{F-move} is always trivial, namely it forms a co-boundary of $A\times K$ group (for fermion SPT phases, the fermion parity $\mathbb{Z}_2^f$ should be included in $K$). Nontrivial ASPTs with pure state correspondence are characterized by nontrivial decoration pattern, not the Berry phases, as the Berry phases become invisible in a density matrix. 

We argue this phase is trivial for the $2d$ SPT phases from decorating $1d$ $K$-SPT on the codimension-1 $A$-defect. From K\"unneth formula, this type of $2d$ SPT phases are labeled by the following 3-cocycle
\begin{align}
\nu_3(h_1,h_2,h_3)\in\mathcal{H}^1\left(A,\mathcal{H}^2\left[K, U(1)\right]\right),
\label{kunneth}
\end{align}
where $h_j=(a_j,k_j)\in A\times K$. Due to the structure in \eqref{kunneth}, by some gauge transformations (attaching 3-coboundaries) we can further write this 3-cocycle in the standard form as
\begin{align}
\nu_3(h_1,h_2,h_3)=[\omega_2(k_1,k_2)](a_3),
\label{3-cocycle}
\end{align}
where $\omega_2(k_1,k_2)$ only depends on $k_1$ and $k_2$ and gives the classification of $1d$ $K$-SPT, and evaluating it on $a_3$ gives the 3-cocycle.


Now we consider the $F$-move of the $A$-defect, see Fig. \ref{F-move}. The $F$-move of the $A$-defect changes the pattern of domain wall configuration of $A$. We consider the Berry phase $\nu_3(\tilde{g_1},\tilde{g_2},\tilde{g_3})=\nu_3(g_0^{-1}g_1,g_1^{-1}g_2,g_2^{-1}g_3)$ that only depends on the group elements of $A$. Due to Eq. \eqref{3-cocycle}, we know the 3-cocycle $\nu_3(\tilde{g_1},\tilde{g_2},\tilde{g_3})$ can be deformed into $\nu_3(1_k,1_k,\tilde{a_3})$, where $k_1=k_2=1_k$ are the identity element of the group $K$, by some gauge transformations. Since the $\nu_3(1_k,1_k,\tilde{a_3})$ now only depends on one group element $\tilde{a_3}$, one can further apply gauge transformation to gauge this phase away as well. As a consequence, all the Berry phase factors can be gauged away for the symmetry group $\mathcal{G}=A\times K$. Therefore, the fidelity strange correlators of ASPT (\ref{SCdef}) and clean SPT (\ref{exact fidelity strange correlator}) are identical for $2d$ SPT from decorating the codimension-1 $A$-defects by $1d$ $K$-SPT. It is easy to generalize our argument to the ASPT phases in higher dimensions with any kind of domain wall decorations, with $\mathcal{G}=A\times K$.

\begin{figure}
    \centering
    \includegraphics[width=0.4\textwidth]{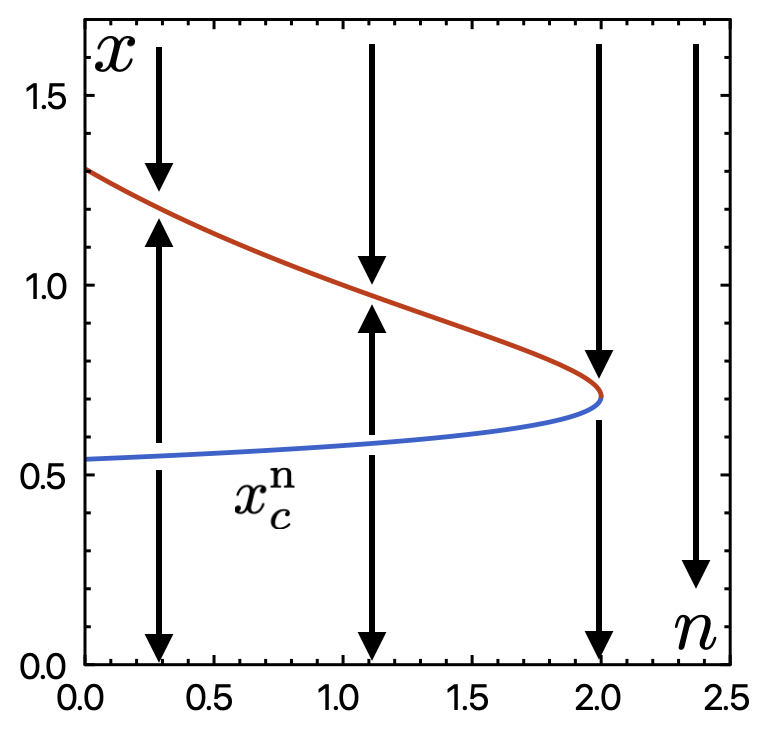}
    \caption{Phase diagram for 2D $O(n)$ loop models.}
    \label{fig:loopPD}
\end{figure}

\section{A review of $O(n)$ loop model and correlation function\label{O(n) loop model}}
In the main text, we have mapped the fidelity strange correlator of $2d$ ASPT phases to a certain correlation function in the self-avoiding $O(n)$ loop model on the honeycomb lattice. Here let us review some of the useful facts on the loop models. The partition function of the loop model is a sum over all possible self-avoiding loop configurations $\mathcal{C}$, weighted by a loop tension $x$ for the total length of loop configurations and a loop fugacity $n$ for the number of loops $|\mathcal{C}|$,
\begin{align}
Z=\sum\limits_{\mathcal{C}}x^{\mathrm{length}}n^{|\mathcal{C}|}.
\end{align}
It is called the $O(n)$ loop model because the partition function of a honeycomb lattice of $O(n)$ spins can be transformed into the above form. 

The phase diagram of the loop model is shown in Fig. \ref{fig:loopPD}. For a fixed value of $n\in[-2,2]$, there is a critical point $x_c=\left[2+\sqrt{2-n}\right]^{-1/2}$ which separates the so-called dense loop phase and dilute loop phase. For $x>x_c$/$x<x_c$, we call the corresponding phase the ``dense/dilute'' loop phase, and $x_c$ remarks their transition termed as the \textit{dilute fixed point} \cite{Duplantier_1989}. 

For the $O(n)$ loop model with $-2\leq n\leq2$, the $L$-leg \textit{watermelon} correlation function $C_L(\bs{x}-\bs{y})$ measures the probability that $L$ non-intersecting lines have a common source point $\bs{x}$ and shrink at the certain endpoint $\bs{y}$. At the dense phase $x>x_c$ and the dilute critical point $x_c$, the $L$-leg watermelon correlation function has power-law decay behavior as
\begin{align}
C_L(\bs{x}-\bs{y})\sim|\bs{x}-\bs{y}|^{-2\Delta_L},
\label{watermelon}
\end{align}
while $C_L(\bs{x}-\bs{y})$ decays exponentially in the dilute loop phase ($x<x_c$). The watermelon correlation is also power law right at the dilute fixed point but with a different exponent for $n<2$.

Let us investigate the critical exponent $\Delta_L$. By the two-dimensional Coulomb gas technique \cite{Nienhuis_book}, the $O(n)$ model can be transformed into a solid-to-solid (SOS) model by orienting the loops in the continuum limit, which is a Gaussian model with the following action
\begin{align}
S_{\mathrm{Gaussian}}=\frac{g}{4\pi}\int\mathrm{d}^2x(\nabla\phi)^2,
\end{align}
where $g$ is the coupling constant of the Coulomb gas, such that
\begin{align}
n=-2\cos(\pi g),
\label{n parameterization}
\end{align}
where $g\in[1,2]$ for the dilute critical point $x_c$ and $g\in[0,1]$ for the dense loop phase ($x>x_c$). The critical exponent $\Delta_L$ of the watermelon correlation function (\ref{watermelon}) is determined by
\begin{align}
\Delta_L=\frac{g}{8}L^2-\frac{(1-g)^2}{2g},
\label{watermelon critical exponent}
\end{align}
for the $O(n)$ model where $n$ is parameterized by Eq. (\ref{n parameterization}). In the main text, we have utilized Eq. (\ref{watermelon}) to calculate various critical exponents of watermelon correlation functions: $L=2$ and $L=3$ for $n=2$ model, and $L=2$ for $n=\sqrt{2}$ model.

\section{Wavefunction overlap of $1d$ dimerized topological states\label{wavefunction overlap}}
In the main text, we emphasize that the wavefunction overlap of the decorated $1d$ SPT phase plays an essential role in the fidelity strange correlator (\ref{SCdef}). Therefore, in this section, we explicitly calculate the wavefunction overlap of $1d$ cluster states with trivial symmetric product state as well as the overlap of $1d$ Kitaev chain with trivial superconductor in systems with finite size using the fixed point wavefunctions. 

For the $1d$ cluster state protected by $\mathbb{Z}_2\times\mathbb{Z}_2$, the wavefunction $|\psi\rangle$ of the $1d$ cluster state in $Z$-basis is explicitly written as
\begin{align}
\langle\cdot\cdot\cdot s_{j-1}s_js_{j+1}\cdot\cdot\cdot|\psi\rangle=(-1)^{\sum_js_js_{j+1}}/\sqrt{2^{2L}}
\label{wavefunction of cluster state}
\end{align}
where $s_j=0,1$ and $2L$ is the number of qubits. Physically, there is a qubit with $X=-1$ at each domain wall of $Z$ (i.e., $Z_{j-1}Z_{j+1}=-1$) and a qubit with $X=1$ at the site away from the $Z$ domain wall. Then consider the trivial state $|\psi_0\rangle$ as the ground state of the disentangled Hamiltonian $H_0=-\sum_jX_j$, with the following explicit form
\begin{align}
|\psi_0\rangle=\bigotimes\limits_{j=1}^{2L}\frac{1}{\sqrt{2}}(|\uparrow\rangle+|\downarrow\rangle)_j=\bigotimes\limits_{j=1}^{2L}|X_j=1\rangle
\label{wavefunction of product state}
\end{align}
Hence the overlap of $1d$ cluster state (\ref{wavefunction of cluster state}) and trivial product state (\ref{wavefunction of product state}) is
\begin{align}
\langle\psi_0|\psi\rangle=\frac{1}{2^{L-1}}=2\times2^{-L}
\end{align}

As we have emphasized in the main text, prefactor 2 is related to the quantum dimension of the boundary of the $1d$ SPT state. 

Next, we consider the Majorana chain, with the following Hamiltonian
\begin{align}
H=-\sum\limits_jc_j^\dag c_{j+1}+h.c.-\sum\limits_{j}c_jc_{j+1}+h.c.
\end{align}
The ground states wavefunctions of the Majorana chain are
\begin{align}
\begin{aligned}
&|\Psi_0^{\mathrm{even}}\rangle=\frac{1}{\sqrt{2}}\left(|\Psi_0^+\rangle+|\Psi_0^-\rangle\right)\\
&|\Psi_0^{\mathrm{odd}}\rangle=\frac{1}{\sqrt{2}}\left(|\Psi_0^+\rangle-|\Psi_0^-\rangle\right)
\end{aligned}
\end{align}
where the superscript even/odd represents the even/odd fermion parity of the ground state wavefunctions, and
\begin{align}
|\Psi_0^\pm\rangle=\frac{1}{2^{L/2}}e^{\pm c_1^\dag}e^{\pm c_2^\dag}\cdot\cdot\cdot e^{\pm c_L^\dag}|0\rangle
\end{align}
It is well-known that the ground state of the Majorana chain with periodic/anti-periodic boundary condition (PBC/anti-PBC) has odd/even fermion parity, hence $|\Psi_0^{\mathrm{even}}$/$|\Psi_0^{\mathrm{odd}}\rangle$ is the ground state wavefunction of the Majorana chain with anti-PBC/PBC. Then consider the atomic insulator as the $1d$ trivial product state, with the following disentangled Hamiltonian
\begin{align}
H_0=\mu\sum\limits_{j}c_j^\dag c_j,~\mu>0
\end{align}
whose ground state wavefunction is the unoccupied vacuum state $|0\rangle$, with even fermion parity. Hence the wavefunction overlap between $|0\rangle$ and $|\Psi_0^{\mathrm{odd}}\rangle$ vanishes because of the different fermion parity, and the wavefunction overlap between $|0\rangle$ and $|\Psi_0^{\mathrm{even}}\rangle$ is
\begin{align}
\langle0|\Psi_0^{\mathrm{even}}\rangle=\left(\frac{1}{\sqrt{2}}\right)^{L-1}=\sqrt{2}\times2^{-L/2},
\end{align}
where the prefactor $\sqrt{2}$ also coincides with the quantum dimension of the boundary Majorana zero modes.

\section{Quantum correction of the loop fugacity}
In the main text, we proved that the fidelity strange correlator of 2$d$ $\Z_2\times\Z_2\times\Z_2^A$ ASPT state is precisely mapped to the loop correlation function in the $O(2)$ loop model. Intuitively, there is only one kind of domain wall of the 2D classical Ising model, hence the Ising domain wall should be described by an $O(1)$ model if we decorate nothing on the Ising domain wall. Here we argue the universal nature of this factor away from the fixed point wavefunction using a field-theoretical representation of the SPT wavefunction. It is known that the wavefunction of a 1d SPT state can be written as $O(3)$ non-linear sigma model (NLSM) with a Wess–Zumino–Witten term at level-1. The overlap between an SPT wavefunction and a trivial state can be represented as \cite{Xu_2013},
\begin{equation}
    \langle\Psi|\Phi\rangle=\int_{\D[\vec{n}(x)]}\exp\left(-\int_{x=0}^L\mathcal{L}[\vec{n}]+i\mathrm{WZW}_1[\vec{n}(x)]\right),
\end{equation}
where $\mathcal{L}[\vec{n}]$ contains the kinetic term of the NLSM~and possible anisotropic terms that break the $SO(3)$ symmetry. Doing a wick rotation, we can equivalently view this overlap as the thermal partition function at temperature $\beta=L$ of a quantum mechanical model of a charged particle moving on a sphere with a $2\pi$ magnetic monopole at the center. This Landau level problem has a robust 2-fold degeneracy of its ground state with energy $\frac{1}{2}\hbar \omega_c$ where $\omega_c$ depends on the details of the kinetic terms. The thermal partition function at large $\beta=L$ is given by $\langle\Psi|\Phi\rangle\cong 2\times e^{-L\hbar\omega_c/2}$, where the prefactor $2$ comes from the 2-fold ground state degeneracy. We also note that this overlap can be viewed as the boundary partition function of the SPT state. Therefore, the degeneracy is also the boundary degeneracy of the SPT that is decorated on the domain wall. 

We can also provide a more rigorous argument based on the matrix product state (MPS) representation of $1d$ SPT phases \cite{chen11a}. A $K$-symmetric $1d$ SPT state can be described by an injective MPS,
\begin{align}
|\Psi_{\mathrm{SPT}}\rangle=\sum\limits_{i_1,\cdot\cdot\cdot i_N}\mathrm{Tr}\left[A_{i_1}\cdot\cdot\cdot A_{i_N}\right]|i_1\cdot\cdot\cdot i_N\rangle,
\end{align}
which has the following symmetry property \cite{chen11a}
\begin{align}
\begin{tikzpicture}[scale=0.99]
\tikzstyle{sergio}=[rectangle,draw=none]
\draw[thick] (0,0) -- (1,0) -- (1,0.6) -- (0,0.6) -- cycle;
\draw[thick] (0,0.3) -- (-0.5,0.3);
\draw[thick] (-1.1,0.3) -- (-1.6,0.3);
\draw[thick] (1,0.3) -- (1.5,0.3);
\draw[thick] (2.1,0.3) -- (2.6,0.3);
\draw[color=blue, ultra thick] (-0.8,0.3) ellipse [x radius=0.3, y radius=0.3];
\draw[color=blue, ultra thick] (1.8,0.3) ellipse [x radius=0.3, y radius=0.3];
\path (-0.8,0.3) node [style=sergio] {$V_g$};
\path (1.8,0.3) node [style=sergio] {$V_g^\dag$};
\path (0.5,0.3) node [style=sergio] {$A_j$};
\draw[thick, densely dashed] (0.5,0.6) -- (0.5,1.1);
\draw[thick] (-4,0) -- (-3,0) -- (-3,0.6) -- (-4,0.6) -- cycle;
\draw[thick] (-4,0.3) -- (-4.5,0.3);
\draw[thick] (-3,0.3) -- (-2.5,0.3);
\path (-3.5,0.3) node [style=sergio] {$A_j$};
\path (-2.1,0.3) node [style=sergio] {$=$};
\draw[thick, densely dashed] (-3.5,0.6) -- (-3.5,1);
\draw[color=blue, ultra thick] (-3.5,1.3) ellipse [x radius=0.3, y radius=0.3];
\path (-3.5,1.3) node [style=sergio] {$U_g$};
\draw[thick, densely dashed] (-3.5,1.6) -- (-3.5,2);
\end{tikzpicture},
\end{align}
where $g\in K$, $U_g$, and $V_g$ are local unitary operators acting on the physical indices (dashed lines) and virtual indices (solid lines), respectively. For $g,h\in K$, the unitary operators $V_g$ and $V_h$ have the following property
\begin{align}
V_gV_h=\omega(g,h)V_{gh},
\end{align}
where $\omega(g,h)\in\mathcal{H}^2[K,U(1)]$ which implies that $V_g$ is a projective representation of the group $K$. 

Then we consider two MPSs of two topologically distinct $1d$ SPT phases $A$ and $B$, which are depicted by two 2-cocycles $\omega_1$ and $\omega_2$ in $\mathcal{H}^2[K,U(1)]$. Their overlap can be represented graphically as
\begin{align}
\begin{tikzpicture}
\tikzstyle{sergio}=[rectangle,draw=none]
\draw[thick] (0,0) -- (1,0) -- (1,0.6) -- (0,0.6) -- cycle;
\draw[thick] (0,0.3) -- (-0.5,0.3);
\draw[thick] (1,0.3) -- (1.5,0.3);
\draw[thick, densely dashed] (0.5,0.6) -- (0.5,1.6);
\draw[thick] (0,1.6) -- (1,1.6) -- (1,2.2) -- (0,2.2) -- cycle;
\draw[thick] (0,0.3+1.6) -- (-0.5,0.3+1.6);
\draw[thick] (1,0.3+1.6) -- (1.5,0.3+1.6);
\draw[thick] (0+2,0) -- (1+2,0) -- (1+2,0.6) -- (0+2,0.6) -- cycle;
\draw[thick] (0+2,0.3) -- (-0.5+2,0.3);
\draw[thick] (1+2,0.3) -- (1.5+2,0.3);
\draw[thick, densely dashed] (0.5+2,0.6) -- (0.5+2,1.6);
\draw[thick] (0+2,1.6) -- (1+2,1.6) -- (1+2,2.2) -- (0+2,2.2) -- cycle;
\draw[thick] (0+2,0.3+1.6) -- (-0.5+2,0.3+1.6);
\draw[thick] (1+2,0.3+1.6) -- (1.5+2,0.3+1.6);
\path (0.5,0.3) node [style=sergio] {$A_j$};
\path (2.5,0.3) node [style=sergio] {$A_{j+1}$};
\path (0.5,0.3+1.6) node [style=sergio] {$B_j$};
\path (2.5,0.3+1.6) node [style=sergio] {$B_{j+1}$};
\path (5.2,1.1) node [style=sergio] {$\cdot\cdot\cdot=\mathrm{Tr}(E_1E_{2}\cdot\cdot\cdot)$};
\path (-1,1.1) node [style=sergio] {$\cdot\cdot\cdot$};
\end{tikzpicture}
\end{align}
where $E_j$ is the transfer matrix of two $1d$ MPSs. Furthermore, by acting $U_g$ and $U_g^\dag$ to the physical indices of two MPSs, we have
\begin{align}
\begin{tikzpicture}[scale=0.9]
\tikzstyle{sergio}=[rectangle,draw=none]
\draw[thick] (0,0) -- (1,0) -- (1,0.6) -- (0,0.6) -- cycle;
\draw[thick] (0,0.3) -- (-0.5,0.3);
\draw[thick] (1,0.3) -- (1.5,0.3);
\draw[thick, densely dashed] (0.5,0.6) -- (0.5,1.6);
\draw[thick] (0,1.6) -- (1,1.6) -- (1,2.2) -- (0,2.2) -- cycle;
\draw[thick] (0,0.3+1.6) -- (-0.5,0.3+1.6);
\draw[thick] (1,0.3+1.6) -- (1.5,0.3+1.6);
\path (0.5,0.3) node [style=sergio] {$A_j$};
\path (0.5,0.3+1.6) node [style=sergio] {$B_j$};
\path (1.75,1.1) node [style=sergio] {$=$};
\draw[thick] (0+2.5,0-0.5) -- (1+2.5,0-0.5) -- (1+2.5,0.6-0.5) -- (0+2.5,0.6-0.5) -- cycle;
\draw[thick] (0+2.5,0.3-0.5) -- (-0.5+2.5,0.3-0.5);
\draw[thick] (1+2.5,0.3-0.5) -- (1.5+2.5,0.3-0.5);
\draw[thick, densely dashed] (0.5+2.5,1.35+0.5) -- (0.5+2.5,1.6+0.5);
\draw[thick, densely dashed] (0.5+2.5,0.1) -- (0.5+2.5,0.35);
\draw[densely dashed] (0.5+2.5,1.15) -- (0.5+2.5,0.95);
\draw[thick] (0+2.5,1.6+0.5) -- (1+2.5,1.6+0.5) -- (1+2.5,2.2+0.5) -- (0+2.5,2.2+0.5) -- cycle;
\draw[thick] (0+2.5,0.3+1.6+0.5) -- (-0.5+2.5,0.3+1.6+0.5);
\draw[thick] (1+2.5,0.3+1.6+0.5) -- (1.5+2.5,0.3+1.6+0.5);
\path (0.5+2.5,0.3-0.5) node [style=sergio] {$A_j$};
\path (0.5+2.5,0.3+1.6+0.5) node [style=sergio] {$B_j$};
\draw[color=blue, ultra thick] (3,1.55) ellipse [x radius=0.3, y radius=0.3];
\path (3,1.55) node [style=sergio] {$U_g^\dag$};
\draw[color=blue, ultra thick] (3,0.65) ellipse [x radius=0.3, y radius=0.3];
\path (3,0.65) node [style=sergio] {$U_g$};
\path (4.25,1.1) node [style=sergio] {$=$};
\draw[thick] (4.5,0.3) -- (4.8,0.3);
\draw[color=blue, ultra thick] (5.1,0.3) ellipse [x radius=0.3, y radius=0.3];
\path (5.1,0.3) node [style=sergio] {$W_g$};
\draw[thick] (5.4,0.3) -- (5.7,0.3);
\draw[thick] (5.7,0) -- (5.7+1,0) -- (5.7+1,0.6) -- (5.7,0.6) -- cycle;
\draw[color=blue, ultra thick] (7.3,0.3) ellipse [x radius=0.3, y radius=0.3];
\draw[thick] (6.7,0.3) -- (7,0.3);
\draw[thick] (7.6,0.3) -- (7.9,0.3);
\path (7.3,0.3) node [style=sergio] {$W_g^\dag$};
\path (6.2,0.3) node [style=sergio] {$A_j$};
\draw[thick, densely dashed] (6.2,0.6) -- (6.2,1.6);
\draw[thick] (4.5,0.3+1.6) -- (4.8,0.3+1.6);
\draw[color=blue, ultra thick] (5.1,0.3+1.6) ellipse [x radius=0.3, y radius=0.3];
\path (5.1,0.3+1.6) node [style=sergio] {$V_g^\dag$};
\draw[thick] (5.4,0.3+1.6) -- (5.7,0.3+1.6);
\draw[thick] (5.7,0+1.6) -- (5.7+1,0+1.6) -- (5.7+1,0.6+1.6) -- (5.7,0.6+1.6) -- cycle;
\draw[color=blue, ultra thick] (7.3,0.3+1.6) ellipse [x radius=0.3, y radius=0.3];
\draw[thick] (6.7,0.3+1.6) -- (7,0.3+1.6);
\draw[thick] (7.6,0.3+1.6) -- (7.9,0.3+1.6);
\path (7.3,0.3+1.6) node [style=sergio] {$V_g$};
\path (6.2,0.3+1.6) node [style=sergio] {$B_j$};
\end{tikzpicture}
\end{align}
i.e., $X_gE_jX_g^\dag=E_j$, where $X_g=V_g^\dag\otimes W_g$ that also satisfies the condition of projective representation of $K$ as
\begin{align}
X_gX_h=\frac{\omega_2(g,h)}{\omega_1(g,h)}X_{gh}
\label{transfer matrix projectively}
\end{align}
Therefore, the transfer matrix $E_j$ has a $K$ symmetry and transforms projectively under $K$. For translational invariant MPSs, the overlap of these MPSs is $\mathrm{Tr}(E^L)$ (where $L$ is the system size), which is determined by the largest eigenvalue of $E$. 

For the $\mathbb{Z}_2$-classified $1d$ SPT phases, the largest eigenvalue of the transfer matrix will always have 2-fold degeneracy from the projectively imposed symmetry $K$, which leads to an overall factor $2$ of the wavefunction overlap of the trivial and nontrivial $1d$ SPT states. We call this universal factor coming from the decorated $1d$ SPT as \textit{quantum correction of the loop fugacity}.

\bibliography{apssamp}

\providecommand{\noopsort}[1]{}\providecommand{\singleletter}[1]{#1}%
\begin{thebibliography}{72}%
\makeatletter
\providecommand \@ifxundefined [1]{%
 \@ifx{#1\undefined}
}%
\providecommand \@ifnum [1]{%
 \ifnum #1\expandafter \@firstoftwo
 \else \expandafter \@secondoftwo
 \fi
}%
\providecommand \@ifx [1]{%
 \ifx #1\expandafter \@firstoftwo
 \else \expandafter \@secondoftwo
 \fi
}%
\providecommand \natexlab [1]{#1}%
\providecommand \enquote  [1]{``#1''}%
\providecommand \bibnamefont  [1]{#1}%
\providecommand \bibfnamefont [1]{#1}%
\providecommand \citenamefont [1]{#1}%
\providecommand \href@noop [0]{\@secondoftwo}%
\providecommand \href [0]{\begingroup \@sanitize@url \@href}%
\providecommand \@href[1]{\@@startlink{#1}\@@href}%
\providecommand \@@href[1]{\endgroup#1\@@endlink}%
\providecommand \@sanitize@url [0]{\catcode `\\12\catcode `\$12\catcode
  `\&12\catcode `\#12\catcode `\^12\catcode `\_12\catcode `\%12\relax}%
\providecommand \@@startlink[1]{}%
\providecommand \@@endlink[0]{}%
\providecommand \url  [0]{\begingroup\@sanitize@url \@url }%
\providecommand \@url [1]{\endgroup\@href {#1}{\urlprefix }}%
\providecommand \urlprefix  [0]{URL }%
\providecommand \Eprint [0]{\href }%
\providecommand \doibase [0]{http://dx.doi.org/}%
\providecommand \selectlanguage [0]{\@gobble}%
\providecommand \bibinfo  [0]{\@secondoftwo}%
\providecommand \bibfield  [0]{\@secondoftwo}%
\providecommand \translation [1]{[#1]}%
\providecommand \BibitemOpen [0]{}%
\providecommand \bibitemStop [0]{}%
\providecommand \bibitemNoStop [0]{.\EOS\space}%
\providecommand \EOS [0]{\spacefactor3000\relax}%
\providecommand \BibitemShut  [1]{\csname bibitem#1\endcsname}%
\let\auto@bib@innerbib\@empty
\bibitem [{\citenamefont {Gu}\ and\ \citenamefont {Wen}(2009)}]{ZCGu2009}%
  \BibitemOpen
  \bibfield  {author} {\bibinfo {author} {\bibfnamefont {Z.-C.}\ \bibnamefont
  {Gu}}\ and\ \bibinfo {author} {\bibfnamefont {X.-G.}\ \bibnamefont {Wen}},\
  }\href {https://journals.aps.org/prb/abstract/10.1103/PhysRevB.80.155131}
  {\bibfield  {journal} {\bibinfo  {journal} {Phys. Rev. B}\ }\textbf {\bibinfo
  {volume} {80}},\ \bibinfo {pages} {155131} (\bibinfo {year}
  {2009})}\BibitemShut {NoStop}%
\bibitem [{\citenamefont {Chen}\ \emph {et~al.}(2011)\citenamefont {Chen},
  \citenamefont {Gu},\ and\ \citenamefont {Wen}}]{chen11a}%
  \BibitemOpen
  \bibfield  {author} {\bibinfo {author} {\bibfnamefont {X.}~\bibnamefont
  {Chen}}, \bibinfo {author} {\bibfnamefont {Z.-C.}\ \bibnamefont {Gu}}, \ and\
  \bibinfo {author} {\bibfnamefont {X.-G.}\ \bibnamefont {Wen}},\ }\href
  {\doibase 10.1103/PhysRevB.83.035107} {\bibfield  {journal} {\bibinfo
  {journal} {Phys. Rev. B}\ }\textbf {\bibinfo {volume} {83}},\ \bibinfo
  {pages} {035107} (\bibinfo {year} {2011})}\BibitemShut {NoStop}%
\bibitem [{\citenamefont {Chen}\ \emph {et~al.}(2012)\citenamefont {Chen},
  \citenamefont {Gu}, \citenamefont {Liu},\ and\ \citenamefont
  {Wen}}]{XieChenScience}%
  \BibitemOpen
  \bibfield  {author} {\bibinfo {author} {\bibfnamefont {X.}~\bibnamefont
  {Chen}}, \bibinfo {author} {\bibfnamefont {Z.-C.}\ \bibnamefont {Gu}},
  \bibinfo {author} {\bibfnamefont {Z.-X.}\ \bibnamefont {Liu}}, \ and\
  \bibinfo {author} {\bibfnamefont {X.-G.}\ \bibnamefont {Wen}},\ }\href
  {https://science.sciencemag.org/content/338/6114/1604} {\bibfield  {journal}
  {\bibinfo  {journal} {Science}\ }\textbf {\bibinfo {volume} {338}},\ \bibinfo
  {pages} {1604} (\bibinfo {year} {2012})}\BibitemShut {NoStop}%
\bibitem [{\citenamefont {Chen}\ \emph {et~al.}(2013)\citenamefont {Chen},
  \citenamefont {Gu}, \citenamefont {Liu},\ and\ \citenamefont
  {Wen}}]{cohomology}%
  \BibitemOpen
  \bibfield  {author} {\bibinfo {author} {\bibfnamefont {X.}~\bibnamefont
  {Chen}}, \bibinfo {author} {\bibfnamefont {Z.-C.}\ \bibnamefont {Gu}},
  \bibinfo {author} {\bibfnamefont {Z.-X.}\ \bibnamefont {Liu}}, \ and\
  \bibinfo {author} {\bibfnamefont {X.-G.}\ \bibnamefont {Wen}},\ }\href
  {https://journals.aps.org/prb/abstract/10.1103/PhysRevB.87.155114} {\bibfield
   {journal} {\bibinfo  {journal} {Phys. Rev. B}\ }\textbf {\bibinfo {volume}
  {87}},\ \bibinfo {pages} {155114} (\bibinfo {year} {2013})}\BibitemShut
  {NoStop}%
\bibitem [{\citenamefont {Senthil}(2015)}]{Senthil_2015}%
  \BibitemOpen
  \bibfield  {author} {\bibinfo {author} {\bibfnamefont {T.}~\bibnamefont
  {Senthil}},\ }\href {\doibase 10.1146/annurev-conmatphys-031214-014740}
  {\bibfield  {journal} {\bibinfo  {journal} {Annu. Rev. Condens. Matter
  Phys.}\ }\textbf {\bibinfo {volume} {6}},\ \bibinfo {pages} {299} (\bibinfo
  {year} {2015})}\BibitemShut {NoStop}%
\bibitem [{\citenamefont {Lu}\ and\ \citenamefont {Vishwanath}(2012)}]{Lu12}%
  \BibitemOpen
  \bibfield  {author} {\bibinfo {author} {\bibfnamefont {Y.-M.}\ \bibnamefont
  {Lu}}\ and\ \bibinfo {author} {\bibfnamefont {A.}~\bibnamefont
  {Vishwanath}},\ }\href {\doibase 10.1103/PhysRevB.86.125119} {\bibfield
  {journal} {\bibinfo  {journal} {Phys. Rev. B}\ }\textbf {\bibinfo {volume}
  {86}},\ \bibinfo {pages} {125119} (\bibinfo {year} {2012})}\BibitemShut
  {NoStop}%
\bibitem [{\citenamefont {Wang}\ and\ \citenamefont {Gu}(2020)}]{general2}%
  \BibitemOpen
  \bibfield  {author} {\bibinfo {author} {\bibfnamefont {Q.-R.}\ \bibnamefont
  {Wang}}\ and\ \bibinfo {author} {\bibfnamefont {Z.-C.}\ \bibnamefont {Gu}},\
  }\href {\doibase 10.1103/PhysRevX.10.031055} {\bibfield  {journal} {\bibinfo
  {journal} {Phys. Rev. X}\ }\textbf {\bibinfo {volume} {10}},\ \bibinfo
  {pages} {031055} (\bibinfo {year} {2020})},\ \Eprint
  {http://arxiv.org/abs/1811.00536} {arXiv:1811.00536 [cond-mat.str-el]}
  \BibitemShut {NoStop}%
\bibitem [{\citenamefont {Kapustin}\ and\ \citenamefont
  {Thorngren}(2017)}]{Kapustin2017}%
  \BibitemOpen
  \bibfield  {author} {\bibinfo {author} {\bibfnamefont {A.}~\bibnamefont
  {Kapustin}}\ and\ \bibinfo {author} {\bibfnamefont {R.}~\bibnamefont
  {Thorngren}},\ }\href {\doibase 10.1007/JHEP10(2017)080} {\bibfield
  {journal} {\bibinfo  {journal} {Journal of High Energy Physics}\ }\textbf
  {\bibinfo {volume} {2017}},\ \bibinfo {pages} {80} (\bibinfo {year}
  {2017})}\BibitemShut {NoStop}%
\bibitem [{\citenamefont {Cheng}\ \emph
  {et~al.}(2018{\natexlab{a}})\citenamefont {Cheng}, \citenamefont {Bi},
  \citenamefont {You},\ and\ \citenamefont {Gu}}]{2DFSPT}%
  \BibitemOpen
  \bibfield  {author} {\bibinfo {author} {\bibfnamefont {M.}~\bibnamefont
  {Cheng}}, \bibinfo {author} {\bibfnamefont {Z.}~\bibnamefont {Bi}}, \bibinfo
  {author} {\bibfnamefont {Y.-Z.}\ \bibnamefont {You}}, \ and\ \bibinfo
  {author} {\bibfnamefont {Z.-C.}\ \bibnamefont {Gu}},\ }\href
  {https://journals.aps.org/prb/abstract/10.1103/PhysRevB.97.205109} {\bibfield
   {journal} {\bibinfo  {journal} {Phys. Rev. B}\ }\textbf {\bibinfo {volume}
  {97}},\ \bibinfo {pages} {205109} (\bibinfo {year}
  {2018}{\natexlab{a}})}\BibitemShut {NoStop}%
\bibitem [{\citenamefont {Levin}\ and\ \citenamefont {Gu}(2012)}]{LevinGu}%
  \BibitemOpen
  \bibfield  {author} {\bibinfo {author} {\bibfnamefont {M.}~\bibnamefont
  {Levin}}\ and\ \bibinfo {author} {\bibfnamefont {Z.-C.}\ \bibnamefont {Gu}},\
  }\href {https://journals.aps.org/prb/abstract/10.1103/PhysRevB.86.115109}
  {\bibfield  {journal} {\bibinfo  {journal} {Phys. Rev. B}\ }\textbf {\bibinfo
  {volume} {86}},\ \bibinfo {pages} {115109} (\bibinfo {year}
  {2012})}\BibitemShut {NoStop}%
\bibitem [{\citenamefont {Cheng}\ \emph
  {et~al.}(2018{\natexlab{b}})\citenamefont {Cheng}, \citenamefont
  {Tantivasadakarn},\ and\ \citenamefont {Wang}}]{braiding}%
  \BibitemOpen
  \bibfield  {author} {\bibinfo {author} {\bibfnamefont {M.}~\bibnamefont
  {Cheng}}, \bibinfo {author} {\bibfnamefont {N.}~\bibnamefont
  {Tantivasadakarn}}, \ and\ \bibinfo {author} {\bibfnamefont {C.}~\bibnamefont
  {Wang}},\ }\href
  {https://journals.aps.org/prx/abstract/10.1103/PhysRevX.8.011054} {\bibfield
  {journal} {\bibinfo  {journal} {Phys. Rev. X}\ }\textbf {\bibinfo {volume}
  {8}},\ \bibinfo {pages} {011054} (\bibinfo {year}
  {2018}{\natexlab{b}})}\BibitemShut {NoStop}%
\bibitem [{\citenamefont {Vishwanath}\ and\ \citenamefont
  {Senthil}(2013)}]{Ashvin2013}%
  \BibitemOpen
  \bibfield  {author} {\bibinfo {author} {\bibfnamefont {A.}~\bibnamefont
  {Vishwanath}}\ and\ \bibinfo {author} {\bibfnamefont {T.}~\bibnamefont
  {Senthil}},\ }\href
  {https://journals.aps.org/prx/abstract/10.1103/PhysRevX.3.011016} {\bibfield
  {journal} {\bibinfo  {journal} {Phys. Rev. X}\ }\textbf {\bibinfo {volume}
  {3}},\ \bibinfo {pages} {011016} (\bibinfo {year} {2013})}\BibitemShut
  {NoStop}%
\bibitem [{\citenamefont {Wang}\ and\ \citenamefont
  {Senthil}(2013)}]{ChongWang2013}%
  \BibitemOpen
  \bibfield  {author} {\bibinfo {author} {\bibfnamefont {C.}~\bibnamefont
  {Wang}}\ and\ \bibinfo {author} {\bibfnamefont {T.}~\bibnamefont {Senthil}},\
  }\href {https://journals.aps.org/prb/abstract/10.1103/PhysRevB.87.235122}
  {\bibfield  {journal} {\bibinfo  {journal} {Phys. Rev. B}\ }\textbf {\bibinfo
  {volume} {87}},\ \bibinfo {pages} {235122} (\bibinfo {year}
  {2013})}\BibitemShut {NoStop}%
\bibitem [{\citenamefont {Wang}\ \emph {et~al.}(2013)\citenamefont {Wang},
  \citenamefont {Potter},\ and\ \citenamefont {Senthil}}]{Senthil2013}%
  \BibitemOpen
  \bibfield  {author} {\bibinfo {author} {\bibfnamefont {C.}~\bibnamefont
  {Wang}}, \bibinfo {author} {\bibfnamefont {A.~C.}\ \bibnamefont {Potter}}, \
  and\ \bibinfo {author} {\bibfnamefont {T.}~\bibnamefont {Senthil}},\ }\href
  {https://journals.aps.org/prb/abstract/10.1103/PhysRevB.88.115137} {\bibfield
   {journal} {\bibinfo  {journal} {Phys. Rev. B}\ }\textbf {\bibinfo {volume}
  {88}},\ \bibinfo {pages} {115137} (\bibinfo {year} {2013})}\BibitemShut
  {NoStop}%
\bibitem [{\citenamefont {Fidkowski}\ \emph {et~al.}(2013)\citenamefont
  {Fidkowski}, \citenamefont {Chen},\ and\ \citenamefont
  {Vishwanath}}]{Lukasz2013}%
  \BibitemOpen
  \bibfield  {author} {\bibinfo {author} {\bibfnamefont {L.}~\bibnamefont
  {Fidkowski}}, \bibinfo {author} {\bibfnamefont {X.}~\bibnamefont {Chen}}, \
  and\ \bibinfo {author} {\bibfnamefont {A.}~\bibnamefont {Vishwanath}},\
  }\href {https://journals.aps.org/prx/abstract/10.1103/PhysRevX.3.041016}
  {\bibfield  {journal} {\bibinfo  {journal} {Phys. Rev. X}\ }\textbf {\bibinfo
  {volume} {3}},\ \bibinfo {pages} {041016} (\bibinfo {year}
  {2013})}\BibitemShut {NoStop}%
\bibitem [{\citenamefont {Wang}\ and\ \citenamefont
  {Senthil}(2014)}]{ChongWang2014}%
  \BibitemOpen
  \bibfield  {author} {\bibinfo {author} {\bibfnamefont {C.}~\bibnamefont
  {Wang}}\ and\ \bibinfo {author} {\bibfnamefont {T.}~\bibnamefont {Senthil}},\
  }\href {\doibase 10.1103/PhysRevB.89.195124} {\bibfield  {journal} {\bibinfo
  {journal} {Phys. Rev. B}\ }\textbf {\bibinfo {volume} {89}},\ \bibinfo
  {pages} {195124} (\bibinfo {year} {2014})}\BibitemShut {NoStop}%
\bibitem [{\citenamefont {Fu}(2011)}]{TCI}%
  \BibitemOpen
  \bibfield  {author} {\bibinfo {author} {\bibfnamefont {L.}~\bibnamefont
  {Fu}},\ }\href
  {https://journals.aps.org/prl/abstract/10.1103/PhysRevLett.106.106802}
  {\bibfield  {journal} {\bibinfo  {journal} {Phys. Rev. Lett.}\ }\textbf
  {\bibinfo {volume} {106}},\ \bibinfo {pages} {106802} (\bibinfo {year}
  {2011})}\BibitemShut {NoStop}%
\bibitem [{\citenamefont {Hsieh}\ \emph {et~al.}(2012)\citenamefont {Hsieh},
  \citenamefont {Lin}, \citenamefont {Liu}, \citenamefont {Duan}, \citenamefont
  {Bansil},\ and\ \citenamefont {Fu}}]{Fu2012}%
  \BibitemOpen
  \bibfield  {author} {\bibinfo {author} {\bibfnamefont {T.~H.}\ \bibnamefont
  {Hsieh}}, \bibinfo {author} {\bibfnamefont {H.}~\bibnamefont {Lin}}, \bibinfo
  {author} {\bibfnamefont {J.}~\bibnamefont {Liu}}, \bibinfo {author}
  {\bibfnamefont {W.}~\bibnamefont {Duan}}, \bibinfo {author} {\bibfnamefont
  {A.}~\bibnamefont {Bansil}}, \ and\ \bibinfo {author} {\bibfnamefont
  {L.}~\bibnamefont {Fu}},\ }\href {\doibase 10.1038/ncomms1969} {\bibfield
  {journal} {\bibinfo  {journal} {Nat. Commun.}\ }\textbf {\bibinfo {volume}
  {3}},\ \bibinfo {pages} {982} (\bibinfo {year} {2012})}\BibitemShut {NoStop}%
\bibitem [{\citenamefont {Isobe}\ and\ \citenamefont {Fu}(2015)}]{ITCI}%
  \BibitemOpen
  \bibfield  {author} {\bibinfo {author} {\bibfnamefont {H.}~\bibnamefont
  {Isobe}}\ and\ \bibinfo {author} {\bibfnamefont {L.}~\bibnamefont {Fu}},\
  }\href {https://journals.aps.org/prb/abstract/10.1103/PhysRevB.92.081304}
  {\bibfield  {journal} {\bibinfo  {journal} {Phys. Rev. B}\ }\textbf {\bibinfo
  {volume} {92}},\ \bibinfo {pages} {081304(R)} (\bibinfo {year}
  {2015})}\BibitemShut {NoStop}%
\bibitem [{\citenamefont {Song}\ \emph {et~al.}(2017)\citenamefont {Song},
  \citenamefont {Huang}, \citenamefont {Fu},\ and\ \citenamefont
  {Hermele}}]{reduction}%
  \BibitemOpen
  \bibfield  {author} {\bibinfo {author} {\bibfnamefont {H.}~\bibnamefont
  {Song}}, \bibinfo {author} {\bibfnamefont {S.-J.}\ \bibnamefont {Huang}},
  \bibinfo {author} {\bibfnamefont {L.}~\bibnamefont {Fu}}, \ and\ \bibinfo
  {author} {\bibfnamefont {M.}~\bibnamefont {Hermele}},\ }\href
  {https://journals.aps.org/prx/abstract/10.1103/PhysRevX.7.011020} {\bibfield
  {journal} {\bibinfo  {journal} {Phys. Rev. X}\ }\textbf {\bibinfo {volume}
  {7}},\ \bibinfo {pages} {011020} (\bibinfo {year} {2017})}\BibitemShut
  {NoStop}%
\bibitem [{\citenamefont {Huang}\ \emph {et~al.}(2017)\citenamefont {Huang},
  \citenamefont {Song}, \citenamefont {Huang},\ and\ \citenamefont
  {Hermele}}]{building}%
  \BibitemOpen
  \bibfield  {author} {\bibinfo {author} {\bibfnamefont {S.-J.}\ \bibnamefont
  {Huang}}, \bibinfo {author} {\bibfnamefont {H.}~\bibnamefont {Song}},
  \bibinfo {author} {\bibfnamefont {Y.-P.}\ \bibnamefont {Huang}}, \ and\
  \bibinfo {author} {\bibfnamefont {M.}~\bibnamefont {Hermele}},\ }\href
  {https://journals.aps.org/prb/abstract/10.1103/PhysRevB.96.205106} {\bibfield
   {journal} {\bibinfo  {journal} {Phys. Rev. B}\ }\textbf {\bibinfo {volume}
  {96}},\ \bibinfo {pages} {205106} (\bibinfo {year} {2017})}\BibitemShut
  {NoStop}%
\bibitem [{\citenamefont {Thorngren}\ and\ \citenamefont
  {Else}(2018)}]{correspondence}%
  \BibitemOpen
  \bibfield  {author} {\bibinfo {author} {\bibfnamefont {R.}~\bibnamefont
  {Thorngren}}\ and\ \bibinfo {author} {\bibfnamefont {D.~V.}\ \bibnamefont
  {Else}},\ }\href
  {https://journals.aps.org/prx/abstract/10.1103/PhysRevX.8.011040} {\bibfield
  {journal} {\bibinfo  {journal} {Phys. Rev. X}\ }\textbf {\bibinfo {volume}
  {8}},\ \bibinfo {pages} {011040} (\bibinfo {year} {2018})}\BibitemShut
  {NoStop}%
\bibitem [{\citenamefont {Po}\ \emph {et~al.}(2017)\citenamefont {Po},
  \citenamefont {Vishwanath},\ and\ \citenamefont {Watanabe}}]{230}%
  \BibitemOpen
  \bibfield  {author} {\bibinfo {author} {\bibfnamefont {H.~C.}\ \bibnamefont
  {Po}}, \bibinfo {author} {\bibfnamefont {A.}~\bibnamefont {Vishwanath}}, \
  and\ \bibinfo {author} {\bibfnamefont {H.}~\bibnamefont {Watanabe}},\ }\href
  {https://www.nature.com/articles/s41467-017-00133-2} {\bibfield  {journal}
  {\bibinfo  {journal} {Nature Communications}\ }\textbf {\bibinfo {volume}
  {8}},\ \bibinfo {pages} {50} (\bibinfo {year} {2017})}\BibitemShut {NoStop}%
\bibitem [{\citenamefont {Cheng}\ and\ \citenamefont {Wang}()}]{rotation}%
  \BibitemOpen
  \bibfield  {author} {\bibinfo {author} {\bibfnamefont {M.}~\bibnamefont
  {Cheng}}\ and\ \bibinfo {author} {\bibfnamefont {C.}~\bibnamefont {Wang}},\
  }\href@noop {} {\ }\Eprint {http://arxiv.org/abs/1810.12308}
  {arXiv:1810.12308 [cond-mat.str-el]} \BibitemShut {NoStop}%
\bibitem [{\citenamefont {Zhang}\ \emph
  {et~al.}(2020{\natexlab{a}})\citenamefont {Zhang}, \citenamefont {Wang},
  \citenamefont {Yang}, \citenamefont {Qi},\ and\ \citenamefont
  {Gu}}]{dihedral}%
  \BibitemOpen
  \bibfield  {author} {\bibinfo {author} {\bibfnamefont {J.-H.}\ \bibnamefont
  {Zhang}}, \bibinfo {author} {\bibfnamefont {Q.-R.}\ \bibnamefont {Wang}},
  \bibinfo {author} {\bibfnamefont {S.}~\bibnamefont {Yang}}, \bibinfo {author}
  {\bibfnamefont {Y.}~\bibnamefont {Qi}}, \ and\ \bibinfo {author}
  {\bibfnamefont {Z.-C.}\ \bibnamefont {Gu}},\ }\href {\doibase
  10.1103/PhysRevB.101.100501} {\bibfield  {journal} {\bibinfo  {journal}
  {Phys. Rev. B}\ }\textbf {\bibinfo {volume} {101}},\ \bibinfo {pages}
  {100501(R)} (\bibinfo {year} {2020}{\natexlab{a}})}\BibitemShut {NoStop}%
\bibitem [{\citenamefont {Zhang}\ \emph
  {et~al.}(2020{\natexlab{b}})\citenamefont {Zhang}, \citenamefont {Yang},
  \citenamefont {Qi},\ and\ \citenamefont {Gu}}]{wallpaper}%
  \BibitemOpen
  \bibfield  {author} {\bibinfo {author} {\bibfnamefont {J.-H.}\ \bibnamefont
  {Zhang}}, \bibinfo {author} {\bibfnamefont {S.}~\bibnamefont {Yang}},
  \bibinfo {author} {\bibfnamefont {Y.}~\bibnamefont {Qi}}, \ and\ \bibinfo
  {author} {\bibfnamefont {Z.-C.}\ \bibnamefont {Gu}},\ }\href {\doibase
  10.1103/PhysRevResearch.4.033081} {\bibfield  {journal} {\bibinfo  {journal}
  {Phys. Rev. Research 4, 033081 (2022)}\ } (\bibinfo {year}
  {2020}{\natexlab{b}}),\ 10.1103/PhysRevResearch.4.033081},\ \Eprint
  {http://arxiv.org/abs/2012.15657} {arXiv:2012.15657 [cond-mat.str-el]}
  \BibitemShut {NoStop}%
\bibitem [{\citenamefont {Tang}\ \emph {et~al.}(2019)\citenamefont {Tang},
  \citenamefont {Po}, \citenamefont {Vishwanath},\ and\ \citenamefont
  {Wan}}]{indicator1}%
  \BibitemOpen
  \bibfield  {author} {\bibinfo {author} {\bibfnamefont {F.}~\bibnamefont
  {Tang}}, \bibinfo {author} {\bibfnamefont {H.~C.}\ \bibnamefont {Po}},
  \bibinfo {author} {\bibfnamefont {A.}~\bibnamefont {Vishwanath}}, \ and\
  \bibinfo {author} {\bibfnamefont {X.}~\bibnamefont {Wan}},\ }\href {\doibase
  10.1038/s41586-019-0937-5} {\bibfield  {journal} {\bibinfo  {journal}
  {Nature}\ }\textbf {\bibinfo {volume} {566}},\ \bibinfo {pages} {486}
  (\bibinfo {year} {2019})}\BibitemShut {NoStop}%
\bibitem [{\citenamefont {Zhang}\ \emph {et~al.}(2022)\citenamefont {Zhang},
  \citenamefont {Qi},\ and\ \citenamefont {Gu}}]{JHZhang2022}%
  \BibitemOpen
  \bibfield  {author} {\bibinfo {author} {\bibfnamefont {J.-H.}\ \bibnamefont
  {Zhang}}, \bibinfo {author} {\bibfnamefont {Y.}~\bibnamefont {Qi}}, \ and\
  \bibinfo {author} {\bibfnamefont {Z.-C.}\ \bibnamefont {Gu}},\ }\href@noop {}
  {\  (\bibinfo {year} {2022})},\ \Eprint {http://arxiv.org/abs/2204.13558}
  {arXiv:2204.13558 [cond-mat.str-el]} \BibitemShut {NoStop}%
\bibitem [{\citenamefont {Zhang}(2022)}]{Zhang_2022}%
  \BibitemOpen
  \bibfield  {author} {\bibinfo {author} {\bibfnamefont {J.-H.}\ \bibnamefont
  {Zhang}},\ }\href {\doibase 10.1103/PhysRevB.106.L020503} {\bibfield
  {journal} {\bibinfo  {journal} {Physical Review B}\ }\textbf {\bibinfo
  {volume} {106}},\ \bibinfo {pages} {l020503} (\bibinfo {year}
  {2022})}\BibitemShut {NoStop}%
\bibitem [{\citenamefont {Kruthoff}\ \emph {et~al.}(2017)\citenamefont
  {Kruthoff}, \citenamefont {de~Boer}, \citenamefont {van Wezel}, \citenamefont
  {Kane},\ and\ \citenamefont {Slager}}]{Kruthoff_2017}%
  \BibitemOpen
  \bibfield  {author} {\bibinfo {author} {\bibfnamefont {J.}~\bibnamefont
  {Kruthoff}}, \bibinfo {author} {\bibfnamefont {J.}~\bibnamefont {de~Boer}},
  \bibinfo {author} {\bibfnamefont {J.}~\bibnamefont {van Wezel}}, \bibinfo
  {author} {\bibfnamefont {C.~L.}\ \bibnamefont {Kane}}, \ and\ \bibinfo
  {author} {\bibfnamefont {R.-J.}\ \bibnamefont {Slager}},\ }\href {\doibase
  10.1103/PhysRevX.7.041069} {\bibfield  {journal} {\bibinfo  {journal}
  {Physical Review X}\ }\textbf {\bibinfo {volume} {7}},\ \bibinfo {pages}
  {041069} (\bibinfo {year} {2017})}\BibitemShut {NoStop}%
\bibitem [{\citenamefont {Slager}\ \emph {et~al.}(2012)\citenamefont {Slager},
  \citenamefont {Mesaros}, \citenamefont {Juri{\v{c}}i{\'{c}}},\ and\
  \citenamefont {Zaanen}}]{Slager_2012}%
  \BibitemOpen
  \bibfield  {author} {\bibinfo {author} {\bibfnamefont {R.-J.}\ \bibnamefont
  {Slager}}, \bibinfo {author} {\bibfnamefont {A.}~\bibnamefont {Mesaros}},
  \bibinfo {author} {\bibfnamefont {V.}~\bibnamefont {Juri{\v{c}}i{\'{c}}}}, \
  and\ \bibinfo {author} {\bibfnamefont {J.}~\bibnamefont {Zaanen}},\ }\href
  {\doibase 10.1038/nphys2513} {\bibfield  {journal} {\bibinfo  {journal}
  {Nature Physics}\ }\textbf {\bibinfo {volume} {9}},\ \bibinfo {pages} {98}
  (\bibinfo {year} {2012})}\BibitemShut {NoStop}%
\bibitem [{\citenamefont {Ma}\ and\ \citenamefont {Wang}(2022)}]{MaWangASPT}%
  \BibitemOpen
  \bibfield  {author} {\bibinfo {author} {\bibfnamefont {R.}~\bibnamefont
  {Ma}}\ and\ \bibinfo {author} {\bibfnamefont {C.}~\bibnamefont {Wang}},\
  }\href@noop {} {\  (\bibinfo {year} {2022})},\ \Eprint
  {http://arxiv.org/abs/2209.02723} {arXiv:2209.02723 [cond-mat.str-el]}
  \BibitemShut {NoStop}%
\bibitem [{\citenamefont {Ringel}\ \emph {et~al.}(2012)\citenamefont {Ringel},
  \citenamefont {Kraus},\ and\ \citenamefont {Stern}}]{Ringel_2012}%
  \BibitemOpen
  \bibfield  {author} {\bibinfo {author} {\bibfnamefont {Z.}~\bibnamefont
  {Ringel}}, \bibinfo {author} {\bibfnamefont {Y.~E.}\ \bibnamefont {Kraus}}, \
  and\ \bibinfo {author} {\bibfnamefont {A.}~\bibnamefont {Stern}},\ }\href
  {\doibase 10.1103/PhysRevB.86.045102} {\bibfield  {journal} {\bibinfo
  {journal} {Physical Review B}\ }\textbf {\bibinfo {volume} {86}},\ \bibinfo
  {pages} {045102} (\bibinfo {year} {2012})}\BibitemShut {NoStop}%
\bibitem [{\citenamefont {Mong}\ \emph {et~al.}(2012)\citenamefont {Mong},
  \citenamefont {Bardarson},\ and\ \citenamefont {Moore}}]{Mong_2012}%
  \BibitemOpen
  \bibfield  {author} {\bibinfo {author} {\bibfnamefont {R.~S.~K.}\
  \bibnamefont {Mong}}, \bibinfo {author} {\bibfnamefont {J.~H.}\ \bibnamefont
  {Bardarson}}, \ and\ \bibinfo {author} {\bibfnamefont {J.~E.}\ \bibnamefont
  {Moore}},\ }\href {\doibase 10.1103/PhysRevLett.108.076804} {\bibfield
  {journal} {\bibinfo  {journal} {Physical Review Letters}\ }\textbf {\bibinfo
  {volume} {108}},\ \bibinfo {pages} {076804} (\bibinfo {year}
  {2012})}\BibitemShut {NoStop}%
\bibitem [{\citenamefont {Fulga}\ \emph {et~al.}(2014)\citenamefont {Fulga},
  \citenamefont {van Heck}, \citenamefont {Edge},\ and\ \citenamefont
  {Akhmerov}}]{Fulga_2014}%
  \BibitemOpen
  \bibfield  {author} {\bibinfo {author} {\bibfnamefont {I.~C.}\ \bibnamefont
  {Fulga}}, \bibinfo {author} {\bibfnamefont {B.}~\bibnamefont {van Heck}},
  \bibinfo {author} {\bibfnamefont {J.~M.}\ \bibnamefont {Edge}}, \ and\
  \bibinfo {author} {\bibfnamefont {A.~R.}\ \bibnamefont {Akhmerov}},\ }\href
  {\doibase 10.1103/PhysRevB.89.155424} {\bibfield  {journal} {\bibinfo
  {journal} {Physical Review B}\ }\textbf {\bibinfo {volume} {89}},\ \bibinfo
  {pages} {155424} (\bibinfo {year} {2014})}\BibitemShut {NoStop}%
\bibitem [{\citenamefont {Chou}\ \emph {et~al.}(2018)\citenamefont {Chou},
  \citenamefont {Nandkishore},\ and\ \citenamefont {Radzihovsky}}]{Chou_2018}%
  \BibitemOpen
  \bibfield  {author} {\bibinfo {author} {\bibfnamefont {Y.-Z.}\ \bibnamefont
  {Chou}}, \bibinfo {author} {\bibfnamefont {R.~M.}\ \bibnamefont
  {Nandkishore}}, \ and\ \bibinfo {author} {\bibfnamefont {L.}~\bibnamefont
  {Radzihovsky}},\ }\href {\doibase 10.1103/PhysRevB.98.054205} {\bibfield
  {journal} {\bibinfo  {journal} {Physical Review B}\ }\textbf {\bibinfo
  {volume} {98}},\ \bibinfo {pages} {054205} (\bibinfo {year}
  {2018})}\BibitemShut {NoStop}%
\bibitem [{\citenamefont {Chou}\ and\ \citenamefont
  {Nandkishore}(2021)}]{Chou_2021}%
  \BibitemOpen
  \bibfield  {author} {\bibinfo {author} {\bibfnamefont {Y.-Z.}\ \bibnamefont
  {Chou}}\ and\ \bibinfo {author} {\bibfnamefont {R.~M.}\ \bibnamefont
  {Nandkishore}},\ }\href {\doibase 10.1103/PhysRevB.103.075120} {\bibfield
  {journal} {\bibinfo  {journal} {Physical Review B}\ }\textbf {\bibinfo
  {volume} {103}},\ \bibinfo {pages} {075120} (\bibinfo {year}
  {2021})}\BibitemShut {NoStop}%
\bibitem [{\citenamefont {Ma}\ \emph {et~al.}(2025)\citenamefont {Ma},
  \citenamefont {Zhang}, \citenamefont {Bi}, \citenamefont {Cheng},\ and\
  \citenamefont {Wang}}]{ma2025topological}%
  \BibitemOpen
  \bibfield  {author} {\bibinfo {author} {\bibfnamefont {R.}~\bibnamefont
  {Ma}}, \bibinfo {author} {\bibfnamefont {J.-H.}\ \bibnamefont {Zhang}},
  \bibinfo {author} {\bibfnamefont {Z.}~\bibnamefont {Bi}}, \bibinfo {author}
  {\bibfnamefont {M.}~\bibnamefont {Cheng}}, \ and\ \bibinfo {author}
  {\bibfnamefont {C.}~\bibnamefont {Wang}},\ }\href
  {https://arxiv.org/abs/2305.16399} {\enquote {\bibinfo {title} {Topological
  phases with average symmetries: the decohered, the disordered, and the
  intrinsic},}\ } (\bibinfo {year} {2025}),\ \Eprint
  {http://arxiv.org/abs/2305.16399} {arXiv:2305.16399 [cond-mat.str-el]}
  \BibitemShut {NoStop}%
\bibitem [{\citenamefont {You}\ \emph {et~al.}(2014)\citenamefont {You},
  \citenamefont {Bi}, \citenamefont {Rasmussen}, \citenamefont {Slagle},\ and\
  \citenamefont {Xu}}]{Strange}%
  \BibitemOpen
  \bibfield  {author} {\bibinfo {author} {\bibfnamefont {Y.-Z.}\ \bibnamefont
  {You}}, \bibinfo {author} {\bibfnamefont {Z.}~\bibnamefont {Bi}}, \bibinfo
  {author} {\bibfnamefont {A.}~\bibnamefont {Rasmussen}}, \bibinfo {author}
  {\bibfnamefont {K.}~\bibnamefont {Slagle}}, \ and\ \bibinfo {author}
  {\bibfnamefont {C.}~\bibnamefont {Xu}},\ }\href {\doibase
  10.1103/PhysRevLett.112.247202} {\bibfield  {journal} {\bibinfo  {journal}
  {Physical Review Letters}\ }\textbf {\bibinfo {volume} {112}},\ \bibinfo
  {pages} {247202} (\bibinfo {year} {2014})}\BibitemShut {NoStop}%
\bibitem [{\citenamefont {Lepori}\ \emph {et~al.}(2022)\citenamefont {Lepori},
  \citenamefont {Burrello}, \citenamefont {Trombettoni},\ and\ \citenamefont
  {Paganelli}}]{Paganelli_2022}%
  \BibitemOpen
  \bibfield  {author} {\bibinfo {author} {\bibfnamefont {L.}~\bibnamefont
  {Lepori}}, \bibinfo {author} {\bibfnamefont {M.}~\bibnamefont {Burrello}},
  \bibinfo {author} {\bibfnamefont {A.}~\bibnamefont {Trombettoni}}, \ and\
  \bibinfo {author} {\bibfnamefont {S.}~\bibnamefont {Paganelli}},\ }\href@noop
  {} {\  (\bibinfo {year} {2022})},\ \Eprint {http://arxiv.org/abs/2209.04283}
  {arXiv:2209.04283 [cond-mat.str-el]} \BibitemShut {NoStop}%
\bibitem [{\citenamefont {Scaffidi}\ and\ \citenamefont
  {Ringel}(2016)}]{Scaffidi_2016}%
  \BibitemOpen
  \bibfield  {author} {\bibinfo {author} {\bibfnamefont {T.}~\bibnamefont
  {Scaffidi}}\ and\ \bibinfo {author} {\bibfnamefont {Z.}~\bibnamefont
  {Ringel}},\ }\href {\doibase 10.1103/PhysRevB.93.115105} {\bibfield
  {journal} {\bibinfo  {journal} {Physical Review B}\ }\textbf {\bibinfo
  {volume} {93}},\ \bibinfo {pages} {115105} (\bibinfo {year}
  {2016})}\BibitemShut {NoStop}%
\bibitem [{\citenamefont {Wu}\ \emph {et~al.}(2015)\citenamefont {Wu},
  \citenamefont {He}, \citenamefont {You}, \citenamefont {Xu}, \citenamefont
  {Meng},\ and\ \citenamefont {Lu}}]{Wu_2015}%
  \BibitemOpen
  \bibfield  {author} {\bibinfo {author} {\bibfnamefont {H.-Q.}\ \bibnamefont
  {Wu}}, \bibinfo {author} {\bibfnamefont {Y.-Y.}\ \bibnamefont {He}}, \bibinfo
  {author} {\bibfnamefont {Y.-Z.}\ \bibnamefont {You}}, \bibinfo {author}
  {\bibfnamefont {C.}~\bibnamefont {Xu}}, \bibinfo {author} {\bibfnamefont
  {Z.~Y.}\ \bibnamefont {Meng}}, \ and\ \bibinfo {author} {\bibfnamefont
  {Z.-Y.}\ \bibnamefont {Lu}},\ }\href {\doibase 10.1103/PhysRevB.92.165123}
  {\bibfield  {journal} {\bibinfo  {journal} {Physical Review B}\ }\textbf
  {\bibinfo {volume} {92}},\ \bibinfo {pages} {165123} (\bibinfo {year}
  {2015})}\BibitemShut {NoStop}%
\bibitem [{\citenamefont {Vanhove}\ \emph {et~al.}(2018)\citenamefont
  {Vanhove}, \citenamefont {Bal}, \citenamefont {Williamson}, \citenamefont
  {Bultinck}, \citenamefont {Haegeman},\ and\ \citenamefont
  {Verstraete}}]{Vanhove_2018}%
  \BibitemOpen
  \bibfield  {author} {\bibinfo {author} {\bibfnamefont {R.}~\bibnamefont
  {Vanhove}}, \bibinfo {author} {\bibfnamefont {M.}~\bibnamefont {Bal}},
  \bibinfo {author} {\bibfnamefont {D.~J.}\ \bibnamefont {Williamson}},
  \bibinfo {author} {\bibfnamefont {N.}~\bibnamefont {Bultinck}}, \bibinfo
  {author} {\bibfnamefont {J.}~\bibnamefont {Haegeman}}, \ and\ \bibinfo
  {author} {\bibfnamefont {F.}~\bibnamefont {Verstraete}},\ }\href {\doibase
  10.1103/PhysRevLett.121.177203} {\bibfield  {journal} {\bibinfo  {journal}
  {Physical Review Letters}\ }\textbf {\bibinfo {volume} {121}},\ \bibinfo
  {pages} {177203} (\bibinfo {year} {2018})}\BibitemShut {NoStop}%
\bibitem [{\citenamefont {Zhou}\ \emph {et~al.}(2022)\citenamefont {Zhou},
  \citenamefont {Li}, \citenamefont {Yan}, \citenamefont {Ye},\ and\
  \citenamefont {Meng}}]{Meng_2022}%
  \BibitemOpen
  \bibfield  {author} {\bibinfo {author} {\bibfnamefont {C.}~\bibnamefont
  {Zhou}}, \bibinfo {author} {\bibfnamefont {M.-Y.}\ \bibnamefont {Li}},
  \bibinfo {author} {\bibfnamefont {Z.}~\bibnamefont {Yan}}, \bibinfo {author}
  {\bibfnamefont {P.}~\bibnamefont {Ye}}, \ and\ \bibinfo {author}
  {\bibfnamefont {Z.~Y.}\ \bibnamefont {Meng}},\ }\href@noop {} {\  (\bibinfo
  {year} {2022})},\ \Eprint {http://arxiv.org/abs/2209.12917} {arXiv:2209.12917
  [cond-mat.str-el]} \BibitemShut {NoStop}%
\bibitem [{\citenamefont {He}\ \emph {et~al.}(2016)\citenamefont {He},
  \citenamefont {Wu}, \citenamefont {You}, \citenamefont {Xu}, \citenamefont
  {Meng},\ and\ \citenamefont {Lu}}]{he2016bona}%
  \BibitemOpen
  \bibfield  {author} {\bibinfo {author} {\bibfnamefont {Y.-Y.}\ \bibnamefont
  {He}}, \bibinfo {author} {\bibfnamefont {H.-Q.}\ \bibnamefont {Wu}}, \bibinfo
  {author} {\bibfnamefont {Y.-Z.}\ \bibnamefont {You}}, \bibinfo {author}
  {\bibfnamefont {C.}~\bibnamefont {Xu}}, \bibinfo {author} {\bibfnamefont
  {Z.~Y.}\ \bibnamefont {Meng}}, \ and\ \bibinfo {author} {\bibfnamefont
  {Z.-Y.}\ \bibnamefont {Lu}},\ }\href {\doibase 10.1103/PhysRevB.93.115150}
  {\bibfield  {journal} {\bibinfo  {journal} {Physical Review B}\ }\textbf
  {\bibinfo {volume} {93}},\ \bibinfo {pages} {115150} (\bibinfo {year}
  {2016})}\BibitemShut {NoStop}%
\bibitem [{\citenamefont {Wierschem}\ and\ \citenamefont
  {Sengupta}(2014{\natexlab{a}})}]{wierschem2014quenching}%
  \BibitemOpen
  \bibfield  {author} {\bibinfo {author} {\bibfnamefont {K.}~\bibnamefont
  {Wierschem}}\ and\ \bibinfo {author} {\bibfnamefont {P.}~\bibnamefont
  {Sengupta}},\ }\href {\doibase 10.1103/PhysRevLett.112.247203} {\bibfield
  {journal} {\bibinfo  {journal} {Physical Review Letters}\ }\textbf {\bibinfo
  {volume} {112}},\ \bibinfo {pages} {247203} (\bibinfo {year}
  {2014}{\natexlab{a}})}\BibitemShut {NoStop}%
\bibitem [{\citenamefont {Vanhove}\ \emph {et~al.}(2022)\citenamefont
  {Vanhove}, \citenamefont {Lootens}, \citenamefont {Tu},\ and\ \citenamefont
  {Verstraete}}]{vanhove2022topological}%
  \BibitemOpen
  \bibfield  {author} {\bibinfo {author} {\bibfnamefont {R.}~\bibnamefont
  {Vanhove}}, \bibinfo {author} {\bibfnamefont {L.}~\bibnamefont {Lootens}},
  \bibinfo {author} {\bibfnamefont {H.-H.}\ \bibnamefont {Tu}}, \ and\ \bibinfo
  {author} {\bibfnamefont {F.}~\bibnamefont {Verstraete}},\ }\href {\doibase
  10.1088/1751-8121/ac68b1} {\bibfield  {journal} {\bibinfo  {journal} {Journal
  of Physics A: Mathematical and Theoretical}\ }\textbf {\bibinfo {volume}
  {55}},\ \bibinfo {pages} {235002} (\bibinfo {year} {2022})}\BibitemShut
  {NoStop}%
\bibitem [{\citenamefont {Wierschem}\ and\ \citenamefont
  {Beach}(2016)}]{wierschem2016detection}%
  \BibitemOpen
  \bibfield  {author} {\bibinfo {author} {\bibfnamefont {K.}~\bibnamefont
  {Wierschem}}\ and\ \bibinfo {author} {\bibfnamefont {K.}~\bibnamefont
  {Beach}},\ }\href {\doibase 10.1103/PhysRevB.93.245141} {\bibfield  {journal}
  {\bibinfo  {journal} {Physical Review B}\ }\textbf {\bibinfo {volume} {93}},\
  \bibinfo {pages} {245141} (\bibinfo {year} {2016})}\BibitemShut {NoStop}%
\bibitem [{\citenamefont {Wierschem}\ and\ \citenamefont
  {Sengupta}(2014{\natexlab{b}})}]{wierschem2014strange}%
  \BibitemOpen
  \bibfield  {author} {\bibinfo {author} {\bibfnamefont {K.}~\bibnamefont
  {Wierschem}}\ and\ \bibinfo {author} {\bibfnamefont {P.}~\bibnamefont
  {Sengupta}},\ }\href {\doibase 10.1103/PhysRevB.90.115157} {\bibfield
  {journal} {\bibinfo  {journal} {Physical Review B}\ }\textbf {\bibinfo
  {volume} {90}},\ \bibinfo {pages} {115157} (\bibinfo {year}
  {2014}{\natexlab{b}})}\BibitemShut {NoStop}%
\bibitem [{\citenamefont {Lessa}\ \emph
  {et~al.}(2024{\natexlab{a}})\citenamefont {Lessa}, \citenamefont {Cheng},\
  and\ \citenamefont {Wang}}]{lessa2024mixed}%
  \BibitemOpen
  \bibfield  {author} {\bibinfo {author} {\bibfnamefont {L.~A.}\ \bibnamefont
  {Lessa}}, \bibinfo {author} {\bibfnamefont {M.}~\bibnamefont {Cheng}}, \ and\
  \bibinfo {author} {\bibfnamefont {C.}~\bibnamefont {Wang}},\ }\href
  {https://arxiv.org/abs/2401.17357} {\enquote {\bibinfo {title} {Mixed-state
  quantum anomaly and multipartite entanglement},}\ } (\bibinfo {year}
  {2024}{\natexlab{a}}),\ \Eprint {http://arxiv.org/abs/2401.17357}
  {arXiv:2401.17357 [cond-mat.str-el]} \BibitemShut {NoStop}%
\bibitem [{\citenamefont {Sang}\ and\ \citenamefont
  {Hsieh}(2024)}]{sang2024stability}%
  \BibitemOpen
  \bibfield  {author} {\bibinfo {author} {\bibfnamefont {S.}~\bibnamefont
  {Sang}}\ and\ \bibinfo {author} {\bibfnamefont {T.~H.}\ \bibnamefont
  {Hsieh}},\ }\href {https://arxiv.org/abs/2404.07251} {\enquote {\bibinfo
  {title} {Stability of mixed-state quantum phases via finite markov length},}\
  } (\bibinfo {year} {2024}),\ \Eprint {http://arxiv.org/abs/2404.07251}
  {arXiv:2404.07251 [quant-ph]} \BibitemShut {NoStop}%
\bibitem [{\citenamefont {Lee}\ \emph {et~al.}(2023)\citenamefont {Lee},
  \citenamefont {Jian},\ and\ \citenamefont {Xu}}]{Lee_2023}%
  \BibitemOpen
  \bibfield  {author} {\bibinfo {author} {\bibfnamefont {J.~Y.}\ \bibnamefont
  {Lee}}, \bibinfo {author} {\bibfnamefont {C.-M.}\ \bibnamefont {Jian}}, \
  and\ \bibinfo {author} {\bibfnamefont {C.}~\bibnamefont {Xu}},\ }\href
  {\doibase 10.1103/prxquantum.4.030317} {\bibfield  {journal} {\bibinfo
  {journal} {PRX Quantum}\ }\textbf {\bibinfo {volume} {4}} (\bibinfo {year}
  {2023}),\ 10.1103/prxquantum.4.030317}\BibitemShut {NoStop}%
\bibitem [{\citenamefont {Lessa}\ \emph
  {et~al.}(2024{\natexlab{b}})\citenamefont {Lessa}, \citenamefont {Ma},
  \citenamefont {Zhang}, \citenamefont {Bi}, \citenamefont {Cheng},\ and\
  \citenamefont {Wang}}]{lessa2024strong}%
  \BibitemOpen
  \bibfield  {author} {\bibinfo {author} {\bibfnamefont {L.~A.}\ \bibnamefont
  {Lessa}}, \bibinfo {author} {\bibfnamefont {R.}~\bibnamefont {Ma}}, \bibinfo
  {author} {\bibfnamefont {J.-H.}\ \bibnamefont {Zhang}}, \bibinfo {author}
  {\bibfnamefont {Z.}~\bibnamefont {Bi}}, \bibinfo {author} {\bibfnamefont
  {M.}~\bibnamefont {Cheng}}, \ and\ \bibinfo {author} {\bibfnamefont
  {C.}~\bibnamefont {Wang}},\ }\href {https://arxiv.org/abs/2405.03639}
  {\enquote {\bibinfo {title} {Strong-to-weak spontaneous symmetry breaking in
  mixed quantum states},}\ } (\bibinfo {year} {2024}{\natexlab{b}}),\ \Eprint
  {http://arxiv.org/abs/2405.03639} {arXiv:2405.03639 [quant-ph]} \BibitemShut
  {NoStop}%
\bibitem [{\citenamefont {Chen}\ \emph {et~al.}(2014)\citenamefont {Chen},
  \citenamefont {Lu},\ and\ \citenamefont {Vishwanath}}]{XieChen_2014}%
  \BibitemOpen
  \bibfield  {author} {\bibinfo {author} {\bibfnamefont {X.}~\bibnamefont
  {Chen}}, \bibinfo {author} {\bibfnamefont {Y.-M.}\ \bibnamefont {Lu}}, \ and\
  \bibinfo {author} {\bibfnamefont {A.}~\bibnamefont {Vishwanath}},\ }\href
  {\doibase 10.1038/ncomms4507} {\bibfield  {journal} {\bibinfo  {journal}
  {Nature Communications}\ }\textbf {\bibinfo {volume} {5}} (\bibinfo {year}
  {2014}),\ 10.1038/ncomms4507}\BibitemShut {NoStop}%
\bibitem [{sup()}]{supplementary}%
  \BibitemOpen
  \href@noop {} {\bibinfo  {journal} {see Supplementary Materials for more
  details}\ }\BibitemShut {NoStop}%
\bibitem [{\citenamefont {Duplantier}(1989)}]{Duplantier_1989}%
  \BibitemOpen
\bibfield  {journal} {  }\bibfield  {author} {\bibinfo {author} {\bibfnamefont
  {B.}~\bibnamefont {Duplantier}},\ }\href {\doibase
  10.1016/0370-1573(89)90042-2} {\bibfield  {journal} {\bibinfo  {journal}
  {Physics Reports}\ }\textbf {\bibinfo {volume} {184}},\ \bibinfo {pages}
  {229} (\bibinfo {year} {1989})}\BibitemShut {NoStop}%
\bibitem [{\citenamefont {Kitaev}(2006)}]{KITAEVanyon}%
  \BibitemOpen
  \bibfield  {author} {\bibinfo {author} {\bibfnamefont {A.}~\bibnamefont
  {Kitaev}},\ }\href {\doibase https://doi.org/10.1016/j.aop.2005.10.005}
  {\bibfield  {journal} {\bibinfo  {journal} {Annals of Physics}\ }\textbf
  {\bibinfo {volume} {321}},\ \bibinfo {pages} {2} (\bibinfo {year} {2006})},\
  \bibinfo {note} {january Special Issue}\BibitemShut {NoStop}%
\bibitem [{\citenamefont {Fidkowski}\ and\ \citenamefont
  {Kitaev}(2011)}]{1Dfermion}%
  \BibitemOpen
  \bibfield  {author} {\bibinfo {author} {\bibfnamefont {L.}~\bibnamefont
  {Fidkowski}}\ and\ \bibinfo {author} {\bibfnamefont {A.}~\bibnamefont
  {Kitaev}},\ }\href
  {https://journals.aps.org/prb/abstract/10.1103/PhysRevB.83.075103} {\bibfield
   {journal} {\bibinfo  {journal} {Phys. Rev. B}\ }\textbf {\bibinfo {volume}
  {83}},\ \bibinfo {pages} {075103} (\bibinfo {year} {2011})}\BibitemShut
  {NoStop}%
\bibitem [{\citenamefont {Wang}\ \emph {et~al.}(2018)\citenamefont {Wang},
  \citenamefont {Ning},\ and\ \citenamefont {Chen}}]{2DTSC}%
  \BibitemOpen
  \bibfield  {author} {\bibinfo {author} {\bibfnamefont {Z.}~\bibnamefont
  {Wang}}, \bibinfo {author} {\bibfnamefont {S.-Q.}\ \bibnamefont {Ning}}, \
  and\ \bibinfo {author} {\bibfnamefont {X.}~\bibnamefont {Chen}},\ }\href
  {\doibase 10.1103/PhysRevB.98.094502} {\bibfield  {journal} {\bibinfo
  {journal} {Phys. Rev. B}\ }\textbf {\bibinfo {volume} {98}},\ \bibinfo
  {pages} {094502} (\bibinfo {year} {2018})}\BibitemShut {NoStop}%
\bibitem [{\citenamefont {Ware}\ \emph {et~al.}(2016)\citenamefont {Ware},
  \citenamefont {Son}, \citenamefont {Cheng}, \citenamefont {Mishmash},
  \citenamefont {Alicea},\ and\ \citenamefont {Bauer}}]{DWfermion1}%
  \BibitemOpen
  \bibfield  {author} {\bibinfo {author} {\bibfnamefont {B.}~\bibnamefont
  {Ware}}, \bibinfo {author} {\bibfnamefont {J.~H.}\ \bibnamefont {Son}},
  \bibinfo {author} {\bibfnamefont {M.}~\bibnamefont {Cheng}}, \bibinfo
  {author} {\bibfnamefont {R.~V.}\ \bibnamefont {Mishmash}}, \bibinfo {author}
  {\bibfnamefont {J.}~\bibnamefont {Alicea}}, \ and\ \bibinfo {author}
  {\bibfnamefont {B.}~\bibnamefont {Bauer}},\ }\href {\doibase
  10.1103/PhysRevB.94.115127} {\bibfield  {journal} {\bibinfo  {journal} {Phys.
  Rev. B}\ }\textbf {\bibinfo {volume} {94}},\ \bibinfo {pages} {115127}
  (\bibinfo {year} {2016})}\BibitemShut {NoStop}%
\bibitem [{\citenamefont {Tarantino}\ and\ \citenamefont
  {Fidkowski}(2016)}]{DWfermion2}%
  \BibitemOpen
  \bibfield  {author} {\bibinfo {author} {\bibfnamefont {N.}~\bibnamefont
  {Tarantino}}\ and\ \bibinfo {author} {\bibfnamefont {L.}~\bibnamefont
  {Fidkowski}},\ }\href {\doibase 10.1103/PhysRevB.94.115115} {\bibfield
  {journal} {\bibinfo  {journal} {Phys. Rev. B}\ }\textbf {\bibinfo {volume}
  {94}},\ \bibinfo {pages} {115115} (\bibinfo {year} {2016})}\BibitemShut
  {NoStop}%
\bibitem [{\citenamefont {Huang}\ \emph {et~al.}(2020)\citenamefont {Huang},
  \citenamefont {Kueng},\ and\ \citenamefont {Preskill}}]{Huang_2020}%
  \BibitemOpen
  \bibfield  {author} {\bibinfo {author} {\bibfnamefont {H.-Y.}\ \bibnamefont
  {Huang}}, \bibinfo {author} {\bibfnamefont {R.}~\bibnamefont {Kueng}}, \ and\
  \bibinfo {author} {\bibfnamefont {J.}~\bibnamefont {Preskill}},\ }\href
  {\doibase 10.1038/s41567-020-0932-7} {\bibfield  {journal} {\bibinfo
  {journal} {Nature Physics}\ }\textbf {\bibinfo {volume} {16}},\ \bibinfo
  {pages} {1050–1057} (\bibinfo {year} {2020})}\BibitemShut {NoStop}%
\bibitem [{\citenamefont {Rath}\ \emph {et~al.}(2021)\citenamefont {Rath},
  \citenamefont {Branciard}, \citenamefont {Minguzzi},\ and\ \citenamefont
  {Vermersch}}]{Rath_2021}%
  \BibitemOpen
  \bibfield  {author} {\bibinfo {author} {\bibfnamefont {A.}~\bibnamefont
  {Rath}}, \bibinfo {author} {\bibfnamefont {C.}~\bibnamefont {Branciard}},
  \bibinfo {author} {\bibfnamefont {A.}~\bibnamefont {Minguzzi}}, \ and\
  \bibinfo {author} {\bibfnamefont {B.}~\bibnamefont {Vermersch}},\ }\href
  {\doibase 10.1103/physrevlett.127.260501} {\bibfield  {journal} {\bibinfo
  {journal} {Physical Review Letters}\ }\textbf {\bibinfo {volume} {127}}
  (\bibinfo {year} {2021}),\ 10.1103/physrevlett.127.260501}\BibitemShut
  {NoStop}%
\bibitem [{\citenamefont {Hu}\ \emph {et~al.}(2023)\citenamefont {Hu},
  \citenamefont {Choi},\ and\ \citenamefont {You}}]{Hu_2023}%
  \BibitemOpen
  \bibfield  {author} {\bibinfo {author} {\bibfnamefont {H.-Y.}\ \bibnamefont
  {Hu}}, \bibinfo {author} {\bibfnamefont {S.}~\bibnamefont {Choi}}, \ and\
  \bibinfo {author} {\bibfnamefont {Y.-Z.}\ \bibnamefont {You}},\ }\href
  {\doibase 10.1103/PhysRevResearch.5.023027} {\bibfield  {journal} {\bibinfo
  {journal} {Phys. Rev. Res.}\ }\textbf {\bibinfo {volume} {5}},\ \bibinfo
  {pages} {023027} (\bibinfo {year} {2023})}\BibitemShut {NoStop}%
\bibitem [{\citenamefont {Miszczak}\ \emph {et~al.}(2008)\citenamefont
  {Miszczak}, \citenamefont {Puchała}, \citenamefont {Horodecki},
  \citenamefont {Uhlmann},\ and\ \citenamefont
  {Życzkowski}}]{miszczak2008sub}%
  \BibitemOpen
  \bibfield  {author} {\bibinfo {author} {\bibfnamefont {J.~A.}\ \bibnamefont
  {Miszczak}}, \bibinfo {author} {\bibfnamefont {Z.}~\bibnamefont {Puchała}},
  \bibinfo {author} {\bibfnamefont {P.}~\bibnamefont {Horodecki}}, \bibinfo
  {author} {\bibfnamefont {A.}~\bibnamefont {Uhlmann}}, \ and\ \bibinfo
  {author} {\bibfnamefont {K.}~\bibnamefont {Życzkowski}},\ }\href
  {https://arxiv.org/abs/0805.2037} {\enquote {\bibinfo {title} {Sub-- and
  super--fidelity as bounds for quantum fidelity},}\ } (\bibinfo {year}
  {2008}),\ \Eprint {http://arxiv.org/abs/0805.2037} {arXiv:0805.2037
  [quant-ph]} \BibitemShut {NoStop}%
\bibitem [{\citenamefont {Zhang}\ \emph {et~al.}(2024)\citenamefont {Zhang},
  \citenamefont {Xu},\ and\ \citenamefont {Xu}}]{zhang2024fluctuation}%
  \BibitemOpen
  \bibfield  {author} {\bibinfo {author} {\bibfnamefont {J.-H.}\ \bibnamefont
  {Zhang}}, \bibinfo {author} {\bibfnamefont {C.}~\bibnamefont {Xu}}, \ and\
  \bibinfo {author} {\bibfnamefont {Y.}~\bibnamefont {Xu}},\ }\href
  {https://arxiv.org/abs/2409.18944} {\enquote {\bibinfo {title}
  {Fluctuation-dissipation theorem and information geometry in open quantum
  systems},}\ } (\bibinfo {year} {2024}),\ \Eprint
  {http://arxiv.org/abs/2409.18944} {arXiv:2409.18944 [quant-ph]} \BibitemShut
  {NoStop}%
\bibitem [{\citenamefont {Lee}\ \emph {et~al.}(2022)\citenamefont {Lee},
  \citenamefont {You},\ and\ \citenamefont {Xu}}]{DSPT}%
  \BibitemOpen
  \bibfield  {author} {\bibinfo {author} {\bibfnamefont {J.~Y.}\ \bibnamefont
  {Lee}}, \bibinfo {author} {\bibfnamefont {Y.-Z.}\ \bibnamefont {You}}, \ and\
  \bibinfo {author} {\bibfnamefont {C.}~\bibnamefont {Xu}},\ }\href@noop {} {\
  (\bibinfo {year} {2022})},\ \Eprint {http://arxiv.org/abs/2210.16323}
  {arXiv:2210.16323 [cond-mat.str-el]} \BibitemShut {NoStop}%
\bibitem [{\citenamefont {Cardy}(2005)}]{Cardy_2005}%
  \BibitemOpen
  \bibfield  {author} {\bibinfo {author} {\bibfnamefont {J.}~\bibnamefont
  {Cardy}},\ }\href {\doibase 10.1016/j.aop.2005.04.001} {\bibfield  {journal}
  {\bibinfo  {journal} {Annals of Physics}\ }\textbf {\bibinfo {volume}
  {318}},\ \bibinfo {pages} {81} (\bibinfo {year} {2005})}\BibitemShut
  {NoStop}%
\bibitem [{\citenamefont {Nahum}\ \emph {et~al.}(2011)\citenamefont {Nahum},
  \citenamefont {Chalker}, \citenamefont {Serna}, \citenamefont
  {Ortu{\~{n}}o},\ and\ \citenamefont {Somoza}}]{Nahum_2011}%
  \BibitemOpen
  \bibfield  {author} {\bibinfo {author} {\bibfnamefont {A.}~\bibnamefont
  {Nahum}}, \bibinfo {author} {\bibfnamefont {J.~T.}\ \bibnamefont {Chalker}},
  \bibinfo {author} {\bibfnamefont {P.}~\bibnamefont {Serna}}, \bibinfo
  {author} {\bibfnamefont {M.}~\bibnamefont {Ortu{\~{n}}o}}, \ and\ \bibinfo
  {author} {\bibfnamefont {A.~M.}\ \bibnamefont {Somoza}},\ }\href {\doibase
  10.1103/PhysRevLett.107.110601} {\bibfield  {journal} {\bibinfo  {journal}
  {Physical Review Letters}\ }\textbf {\bibinfo {volume} {107}},\ \bibinfo
  {pages} {110601} (\bibinfo {year} {2011})}\BibitemShut {NoStop}%
\bibitem [{\citenamefont {Nahum}(2015)}]{Nahum_2015}%
  \BibitemOpen
  \bibfield  {author} {\bibinfo {author} {\bibfnamefont {A.}~\bibnamefont
  {Nahum}},\ }\href {\doibase 10.1007/978-3-319-06407-9} {\emph {\bibinfo
  {title} {Critical Phenomena in Loop Models}}}\ (\bibinfo  {publisher}
  {Springer International Publishing},\ \bibinfo {year} {2015})\BibitemShut
  {NoStop}%
\bibitem [{\citenamefont {Nienhuis}(1987)}]{Nienhuis_book}%
  \BibitemOpen
  \bibfield  {author} {\bibinfo {author} {\bibfnamefont {B.}~\bibnamefont
  {Nienhuis}},\ }\href@noop {} {\emph {\bibinfo {title} {Phase Transitions and
  Critical Phenomena}}},\ edited by\ \bibinfo {editor} {\bibfnamefont
  {C.}~\bibnamefont {Domb}}\ and\ \bibinfo {editor} {\bibfnamefont {J.~L.}\
  \bibnamefont {Lebowitz}}\ (\bibinfo  {publisher} {Academic, London},\
  \bibinfo {year} {1987})\BibitemShut {NoStop}%
\bibitem [{\citenamefont {Xu}\ and\ \citenamefont {Senthil}(2013)}]{Xu_2013}%
  \BibitemOpen
  \bibfield  {author} {\bibinfo {author} {\bibfnamefont {C.}~\bibnamefont
  {Xu}}\ and\ \bibinfo {author} {\bibfnamefont {T.}~\bibnamefont {Senthil}},\
  }\href {\doibase 10.1103/PhysRevB.87.174412} {\bibfield  {journal} {\bibinfo
  {journal} {Physical Review B}\ }\textbf {\bibinfo {volume} {87}},\ \bibinfo
  {pages} {174412} (\bibinfo {year} {2013})}\BibitemShut {NoStop}%
\end{thebibliography}%

\end{document}